\documentclass[%
 prx,
 amsmath,amssymb,
 reprint,
 superscriptaddress,
 floatfix
]{revtex4-2}

\usepackage{graphicx}
\usepackage{dcolumn}
\usepackage{comment}
\usepackage{bm}
\usepackage{tikz}
\usepackage[breaklinks,colorlinks,bookmarks=false,citecolor=blue,linkcolor=red,urlcolor=blue]{hyperref}
\usepackage{braket}
\DeclareMathOperator{\Rea}{Re}

\newcommand\td{t^\dagger}

\newcommand\btd{\mathbf{t}^\dagger}
\newcommand\bt{\mathbf{t}}
\newcommand\Jph{\hat{\mathcal{J}}}

\newcommand\Safm{S^{\text{AFM}}}
\newcommand\chiafm{\chi^{\text{AFM}}}

\newcommand\JD{J_\text{D}}
\newcommand\mnp[1]{MnP#1\textsubscript{3}}

\newcommand{\normord}[1]{:\mathrel{#1}:}
\newcommand{\bphi}{\boldsymbol{\phi}}
\newcommand{\bpi}{\boldsymbol{\pi}}

\begin{document}
\title{Cavity-renormalized quantum criticality in a honeycomb bilayer antiferromagnet}

\author{Lukas Weber}
\email{lweber@flatironinstitute.org}
\affiliation{Center for Computational Quantum Physics, The Flatiron Institute, 162 Fifth Avenue, New York, New York 10010, USA}
\affiliation{Max Planck Institute for the Structure and Dynamics of Matter,
Luruper Chaussee 149, 22761 Hamburg, Germany}
\author{Emil Vi{\~n}as Bostr{\"o}m}
\affiliation{Max Planck Institute for the Structure and Dynamics of Matter,
Luruper Chaussee 149, 22761 Hamburg, Germany}
\author{Martin Claassen}
\affiliation{Department of Physics and Astronomy, University of Pennsylvania, Philadelphia,
PA 19104, USA}
\author{Angel Rubio}
\affiliation{Max Planck Institute for the Structure and Dynamics of Matter,
Luruper Chaussee 149, 22761 Hamburg, Germany}
\affiliation{Center for Computational Quantum Physics, The Flatiron Institute, 162 Fifth Avenue, New York, New York 10010, USA}
\author{Dante M. Kennes}
\email{dante.kennes@mpsd.mpg.de}
\affiliation{Institute for Theoretical Solid State Physics, RWTH Aachen University, 52062 Aachen, Germany}
\affiliation{JARA-Fundamentals of Future Information Technology, Jülich, Germany}
\affiliation{Max Planck Institute for the Structure and Dynamics of Matter,
Luruper Chaussee 149, 22761 Hamburg, Germany}

\begin{abstract}
Strong light-matter interactions as realized in an optical cavity provide a tantalizing opportunity to control the properties of condensed matter systems. Inspired by experimental advances in cavity quantum electrodynamics and the fabrication and control of two-dimensional magnets, we investigate the fate of a quantum critical antiferromagnet coupled to an optical cavity field. Using unbiased quantum Monte Carlo simulations, we compute the scaling behavior of the magnetic structure factor and other observables. While the position and universality class are not changed by a single cavity mode, the critical fluctuations themselves obtain a sizable enhancement, scaling with a fractional exponent that defies expectations based on simple perturbation theory. The scaling exponent can be understood using a generic scaling argument, based on which we predict that the effect may be even stronger in other universality classes.
Our microscopic model is based on realistic parameters for two-dimensional magnetic quantum materials and the effect may be within the range of experimental detection.
\end{abstract}
\maketitle

\section{Introduction}
Recently, driving quantum systems with light has emerged as an intriguing route for material control. In the case of classical light this amounts to a non-equilibrium problem~\cite{basov_towards_2017}, and when the magnitude of the external drive is strong enough the field can have a profound impact on the matter degrees of freedom. This has led to many ground-breaking results in the field of polaritonic chemistry and beyond \cite{karzig_topological_2015,claassen_universal_2019,basov_polariton_2021,disa_engineering_2021,bloch_strongly_2022}.

Advances in realizing ultra-strong light-matter coupling in optical cavities~\cite{Kockum2019,Hubener2021,Schlawin2022} have paved the way for an alternative approach, where the quantum fluctuations of light are harnessed in an equilibrium setting. In particular, the fluctuations of the electromagnetic modes can couple strongly to the matter and be used to control chemistry~\cite{Hutchison2012,Galego2015,Ebbesen2016,Feist2018,fribeiro_polariton_2018,schafer_modification_2019,sidler_polaritonic_2021,Sidler2022,li_molecular_2022} and material properties~\cite{Hubener2021,genet_inducing_2021,Schlawin2022}. In condensed matter systems, cavities hold the promise of circumventing the heating problems inherent to laser-driving~\cite{DAlessio2014,Lazarides2014,Kennes2018} while achieving similar control over material properties \cite{Sentef2020,Hubener2021}. This includes proposals to realize quantum-light-induced topological phase transitions~\cite{Kibis2011,Sentef2019,Dmytruk2022}, ferro-electricity~\cite{Ashida2020,Latini2021}, excitonic insulators~\cite{Mazza2019}, magnetic phase transitions and quantum spin liquids~\cite{Bostrom2022,Chiocchetta2021} and superconductivity~\cite{Laussy2010,Laplace2016,Sentef2018, Hagenmuller2019,Gao2020,Chakraborty2021}.

In addition, the effect of light on quantum phase transitions and their critical phenomena is of particular interest. Here, the ground state of the system becomes extremely susceptible to external influences~\cite{sachdev_quantum_2011}, so that even a small light-matter coupling to the collective degrees of freedom could have a significant impact. The origin of this susceptibility is the divergence of quantum fluctuations with system size, which also makes quantum critical points prime examples of strongly correlated physics devoid of simple quasi-particle descriptions. The effects of such strongly correlated quantum fluctuations have so far only been scarcely explored in the cavity setting. Understanding them poses the combined challenge of treating quantum many-body systems and the intricacies that arise in low-energy formulations of quantum electrodynamics (QED) in a cavity~\cite{Ruggenthaler2018,rokaj_lightmatter_2018,schafer_relevance_2020,Sidler2022,Ruggenthaler2022}.

The development of numerical methods to treat cavity systems has seen some recent progress, especially in the field of quantum chemistry where the quantum electrodynamical density functional theory (QEDFT)~\cite{Ruggenthaler2014,Flick2017} and coupled cluster theory~\cite{haugland_coupled_2020,mordovina_polaritonic_2020,pavosevic_polaritonic_2021} allows for an accurate {\it ab initio} treatment of molecules in a cavity. Established numerical methods for strongly correlated lattice models, capable of simulating a quantum critical systems, have on the other hand seen little development. Until now, mainly exact diagonalization (ED)~\cite{Rohn2020,Schuler2020} and density matrix renormalization group~\cite{MendezCordoba2020,Chiriaco2022,Passetti2022} studies have been performed, while higher-dimensional tensor-network methods have not yet been applied to the cavity problem. These approaches are either restricted to small systems, quasi-one-dimensional systems, or low entanglement, respectively. This leaves a blind spot for two-dimensional (2D) materials~\cite{novoselov_2d_2016}, which due to their tunability and richness in quantum critical phenomena may be useful platforms to investigate the effects of quantum light on quantum criticality.

\begin{figure}
	\includegraphics{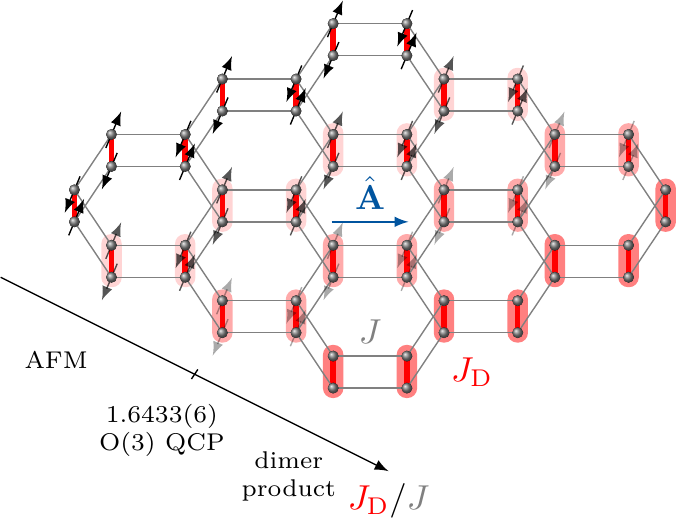}
	\caption{{\bf Magnetic phases of the 2D antiferromagnetic Heisenberg model on a bilayer honeycomb lattice.} Depending on the ratio between the interlayer coupling $\JD$ (bold, red) and intralayer coupling $J$ (thin, gray), the magnetic ground state either forms a N\'eel-type antiferromagnetic order or interlayer singlet dimers, breaking no symmetries. At the phase boundary there is a quantum critical point of the three-dimensional $O(3)$ universality class. Here we consider $\JD\approx \JD^c$ and a coupling to a cavity mode described by the quantum vector potential $\hat{\mathbf{A}}$, linearly polarized along one of the in-plane bond directions.}
	\label{fig:lattice}
\end{figure}

In this work, we address this open issue by presenting a method capable of studying a 2D quantum critical magnet coupled to a single effective cavity mode. Inspired by recent advances in realizing magnetic van der Waals (vdW) materials of atomic thickness~\cite{wang_magnetic_2022,rahman_recent_2021}, as evidenced in particular by the transition metal phosphoruos trichalcogenides $M$P$X_3$ (with $M =$ Fe, Mn or Cr and $X =$ S, Se or Te)~\cite{wildes_spin_1998,chittari_electronic_2016,calder_magnetic_2021}, we consider a Heisenberg-type antiferromagnet (AFM) on a honeycomb bilayer~(Fig.~\ref{fig:lattice}). This system is well known to have a quantum critical point (QCP) in the (2+1)D O(3) universality class at a given ratio of the intra- and interlayer exchange couplings~\cite{millis_spin_1993,chubukov_phase_1995,sandvik_order-disorder_1994,Ganesh2011} at the border between a Néel-ordered AFM state and a quantum-disordered interlayer dimer singlet state. This critical point can be reached by applying hydrostatic pressure, as recently demonstrated for a different magnetic phase transition in CrI\textsubscript{3}~\cite{song_switching_2019}. Furthermore, in cases where the magnetic point group breaks inversion symmetry, the AFM order parameter is accessible via the linear dichroism~\cite{saidl_optical_2017}, reflectance anisotropy~\cite{grigorev_optical_2021}, and via Raman scattering~\cite{kim_antiferromagnetic_2019-1}.

Coupling the magnetic system to cavity photons will influence the spin exchange interactions along the direction of the photon polarization and potentially the quantum phase transition. Our numerical tool to address this question is quantum Monte Carlo simulations, which so far have not seen much use in cavity-matter systems (although spin-boson models in general have been studied~\cite{Sandvik1997,MWeber2022,MWeber2022b}). We find a relevant parameter region where the simulations are sign-problem free, and via simulations of large-size systems reveal that for a single cavity mode the QCP is not shifted. However, the magnetic fluctuations at the critical point experience an enhancement that can be understood as a finite-size correction to scaling, with a small universal fractional scaling exponent that is in stark contrast to the analytic scaling one would expect from simple perturbative arguments in the light-matter coupling. 

The light-induced correction to scaling, while unable to change the universality class of the transition, manifests in an absolute enhancement of the AFM structure factor that remains in the thermodynamic limit, even though the energy content of the single cavity mode remains microscopic. This result may be interpreted as a light-induced change in the ground state of a quantum many-body system.

The remainder of this article is structured as follows. In Sec.~\ref{sec:model} we give a detailed overview of the microscopic model we consider. In Sec.~\ref{sec:qmc} we introduce a quantum Monte Carlo method capable of simulating the light-matter interactions in a magnet in an unbiased way, and present numerical results revealing the fate of the quantum critical point. In Sec.~\ref{sec:fields} we generalize our findings to a larger class on system based on analytic scaling arguments from a continuum field theory. Finally, we draw conclusions in Sec.~\ref{sec:conclusion}.


\section{Cavity-coupled antiferromagnets}
\label{sec:model}
In this section, we introduce the model for the light-matter-coupled antiferromagnet that we will consider. While one possible starting point is to write a phenomenologically motivated light-spin interaction, such an interaction may be missing higher-order terms that are important for the boundedness of the Hamiltonian~\cite{Dmytruk2021,Eckhardt2022}. Therefore, we start instead from a lattice model that is manifestly gauge invariant, the Hubbard model with the Peierls substitution,
\begin{align}
    \label{eq:hubbard}
	H &= \sum_{\braket{ij}\sigma} \big( -t_{ij} e^{i \theta_{ij}} c_{i\sigma}^\dagger c_{j\sigma} +\text{h.c.} \big) \nonumber\\
	&+ U \sum_i n_{i,\downarrow} n_{i,\uparrow} + \Omega a^\dagger a,
\end{align}
where we assume a single relevant effective cavity mode at frequency $\Omega$ in the dipole approximation $\theta_{ij} = (e/\hbar) \int_i^j d\mathbf{r} \cdot \hat{\mathbf{A}} \approx \lambda_{ij} (a^\dagger + a)/\sqrt{N}$. In the large-$U$ limit and at half filling, this model can be down-folded to a Heisenberg-like effective Hamiltonian using a perturbative expansion in $t/U$~\cite{Sentef2020}, resulting in
\begin{align}
	\label{eq:spinham}
	H_\text{eff} = \sum_{\braket{ij}} \Jph_{ij}(a^\dagger, a) \left(\mathbf{S}_i\cdot\mathbf{S}_j - \frac 1 4\right) + \Omega a^\dagger a.
\end{align}
The most striking difference of this Hamiltonian to the regular Heisenberg model is the photon-dependent exchange coupling $\Jph_{ij}(a^\dagger,a)$, which encodes the creation and annihilation of photons during the virtual hopping processes of the electrons mediated by the cavity mode.

The exact form of the downfolded Peierls coupling is quite complex and naive perturbative expansions in $\lambda_{ij}/\sqrt{N}$ can, like in the regular Peierls substitution, lead to unphysical consequences~\cite{Dmytruk2021,Eckhardt2022}. Therefore, we will avoid further approximations and treat the full coupling, $\Jph_{ij}$, exactly. Despite this, the perturbative downfolding itself is not gauge invariant since it hosts a spurious superradiant phase at sufficiently large $\lambda$ (see App.~\ref{app:superrad}). Thus, to remain in the regime of validity of the downfolding, in the remainder of this work, we will restrict ourselves to values of $\lambda$ where the photon occupation remains small and finite in the thermodynamic limit. 

In the following, it is most convenient to express $\Jph_{ij}$ in the occupation-number basis,
\begin{align}
	\label{eq:downfolded_peierls}
	\braket{n|\Jph_{ij}|m} &= \frac{J_{ij}}{2} \sum_{l=0}^\infty \Rea D^{ij}_{nl} (D^{ij}_{lm})^*\nonumber\\
	&\times \left(\frac{1}{1 + \bar{\omega} (l - n)} + \frac{1}{1 + \bar{\omega} (l-m)}\right),
\end{align}
in terms of the normal exchange coupling $J_{ij} = 4t_{ij}^2/U$, the reduced frequency $\bar{\omega} = \Omega/U$ and the displacement operators
\begin{align}
	D^{ij}_{nm} &= \braket{n|e^{i (\lambda_{ij}/\sqrt{N}) (a^\dagger + a)}|m}\nonumber\\
	&= \sqrt{n!m!} \left(\frac{i\lambda_{ij}}{\sqrt{N}}\right)^\delta \sum^{\mu}_{k=0} \frac{e^{-\lambda_{ij}^2/2N}(-\lambda_{ij}^2/N)^k}{k!(\mu-k)!(\delta+k)!},\label{eq:dispsum}
\end{align}
with $\mu = \min\{n,m\}$ and $\delta = |m-n|$.

This expression for the coupling has two key features. First, the even and odd photon number sectors decouple, due to parity conservation. Second, singularities appear when $n\bar{\omega} = 1$ that are associated with degeneracies between photon and doublon electronic excitations. At these singularities, our perturbation theory is expected to break down leading to a different effective model~\cite{Kiffner2019}. In App.~\ref{app:ed_comparison}, we investigate this issue further by comparing our results for a small system to ED results for the Hubbard model.

Considering the large $U/J_{ij}$ regime, where higher-order terms are partly suppressed, one way to maximize the effect of the light-matter coupling is to tune $\bar{\omega}$ close but not too close to one of the singularities. There is, however, a trade-off as high cavity frequencies make cavity excitations less relevant in the ground state, and $n$-photon processes are suppressed by powers of $(\lambda_{ij}/\sqrt{N})^n$ for $n\ge 2$. We find that $\bar{\omega}\approx 1/2$ is a good compromise.

Inspired by \mnp{Se}, we consider the Hamiltonian of Eq.~\eqref{eq:spinham} on the AA-stacked honeycomb bilayer (Fig.~\ref{fig:lattice}). We assume antiferromagnetic exchange couplings both along the nearest-neighbor intralayer bonds, $J$, and the interlayer bonds, $\JD$. The polarization of the cavity mode is chosen so that it aligns with one of the $J$ bonds, compatible with a vanishing in-plane momentum. In this way, it decouples from the $\JD$ bonds, directly influencing the ratio $\JD/J$ that is the relevant coupling at the critical point. Although the magnetic moments in \mnp{Se} are $S=5/2$ and those in our model are $S=1/2$, the Néel-dimer singlet QCP is expected to exist also at higher spin magnitudes~\cite{Ganesh2011}.


\section{Quantum Monte Carlo}
\label{sec:qmc}
To achieve an accurate description of the physics close to the quantum critical point, it is crucial to solve the Hamiltonian of Eq. \eqref{eq:spinham} taking all correlations into account. Without a cavity this is routinely accomplished for unfrustrated quantum magnets using large-scale quantum Monte Carlo simulations in the stochastic series expansion formalism~\cite{Sandvik1999, Syljuasen2002}, which we will outline in the context of our work in the following.

\subsection{Method}
We here extend the stochastic series expansion method to magnets coupled to cavity modes, as exemplified by the down-folded Peierls interaction presented in Sec.~\ref{sec:model}. However, our method applies to any spin-photon Hamiltonian of the form 
\begin{align}
 H_{\rm s-ph} = \sum_{nm} \big( H_{{\rm s},nm} + \Omega n \delta_{nm} \big),
\end{align}
where $H_{{\rm s},nm}$ is a spin Hamiltonian whose parameters are determined by the photon number sector $(nm)$. To apply the stochastic series expansion method, one needs two ingredients: a computational basis and a decomposition of the Hamiltonian into bond terms. Our computational basis is the exterior product of the photon occupation number ($|n\rangle$) and spin-$S_z$ ($|{\uparrow}\rangle$ and $|{\downarrow}\rangle$) bases, where we truncate the photonic Hilbert space at a sufficiently large maximum occupation number, $n < n_{\rm ph}^{\max}$, to achieve converged results. Then, we decompose the Hamiltonian into “three-site” bond operators,
\begin{align}
	H &= \sum_{\braket{ij}} h_{0,ij},\\
	h_{0,ij} &= \Jph_{ij} \left(\mathbf{S}_i \cdot \mathbf{S}_j - \frac 1 4\right) + \frac{\Omega}{N_b} a^\dagger a, 
\end{align}
all acting, apart from regular spin lattice sites $i$ and $j$, on the same artificial “cavity site” denoted “0”, containing the photonic Hilbert space. The $\Omega a^\dagger a$ term is split up evenly and distributed among all $N_b$ bond terms.

In practice, a major obstacle to such extensions is that introducing new couplings to the Hamiltonian can cause the emergence of a sign problem~\cite{Pan2022}. The sign problem leads to an increase in statistical errors that typically fatally decreases the efficiency of the method.

Fortunately, while the addition of the down-folded Peierls coupling does in general cause a sign problem, the model can be made completely sign-problem-free for a large range of parameters using two basic unitary transformations. The first one is a $\pi$-rotation of the spins on one sublattice, mapping $S_i^+S_j^-\mapsto - S_i^+S_j^-$ for all bonds $(ij)$. This transformation is routinely used to make bipartite AFMs sign-free in the $S^z$ basis by making the off-diagonal spin interactions ferromagnetic. In the presence of the cavity, this step alone is not enough since each of the matrix elements of the cavity coupling $\Jph_{ij}$ can add additional signs.

\begin{figure}
	\includegraphics{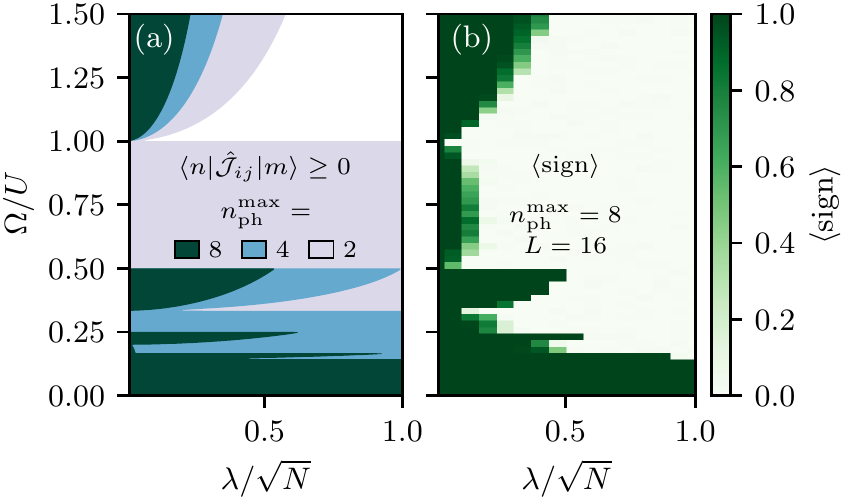}
	\caption{{\bf The sign problem for a coupled spin-photon system.} (a) Exactly sign-free regions fulfilling the condition $\braket{n|\Jph_{ij}|m}\ge 0$ for all $n, m < n_\text{ph}^\text{max}$. (b) Actual average sign in a simulation at $T = J/2L$. For for an average sign $\braket{\text{sign}} = 1$, i.e. outside of the white regions,  efficient large-scale simulations are possible.}
	\label{fig:sign}
\end{figure}

A sufficient (but not necessary) condition for sign-freeness is $\braket{n|\Jph_{ij}|m} \ge 0$ for all $m, n < n_{\rm ph}^\text{max}$. This condition can be fulfilled for a $n_{\rm ph}^\text{max}$-dependent region in parameter space after performing a second, diagonal unitary transformation that maps $a\mapsto i a$. Under this transformation, the matrix elements of the displacement operator, $D_{nm}^{ij}$, become positive, leaving only the signs from the denominators in Eq.~\eqref{eq:downfolded_peierls}. The sum over these denominators is positive in the blue-detuned region of $\bar{\omega} = 1$ and the red-detuned regions of the other singularities $n \bar{\omega} = 1$, as long as $\lambda_{ij}/\sqrt{N}$ is not too large (Fig.~\ref{fig:sign}(a)).
The resulting exactly sign-free regions shrink with increasing photon number cutoff $n_\text{ph}^{\text{max}}$ but grow with system size (due to the factor $1/\sqrt{N}$ in the coupling), so that simulations converged in both the cutoff and system size are possible in these regions.

In our simulations, we find that even in the parameter regions outside of the ones in Fig.~\ref{fig:sign}(a), the sign-problem can be relatively benign at weak coupling $\lambda/\sqrt{N}$ or high $\Omega/U$, where problematic negative matrix elements become very rare in the sampling (Fig.~\ref{fig:sign}(b)).

With all ingredients of a sign-problem-free stochastic series expansion  in place, we use the recently developed abstract loop update algorithm~\cite{Weber2021} to perform QMC sampling in the given basis and bond-operator decomposition without the need of engineering model specific loop update rules. To solve the linear-programming problem that appears when finding the optimal loop propagation probabilities~\cite{Alet2003, Alet2005}, we employ the HiGHs package~\cite{Huangfu2018}.

At this point it is helpful to discuss parallels with the mathematically similar spin-phonon and one-dimensional electron-phonon models, where other stochastic series expansion methods have been developed. While earlier studies relied on rather inefficient local updates~\cite{Sandvik1997,Hardikar2007}, recent advances in the sampling of retarded interactions allow efficient treatment of models where the phonons can be integrated out exactly~\cite{MWeber2022,MWeber2022b}.

Carrying over these advances into the photonic setting is in our case not straightforward due to the highly nonlinear nature of the downfolded Peierls coupling preventing the exact integration of the photons. On the other hand, we note that our method provides a global update for generic nonlinear spin-boson interactions and may in turn be useful in the phononic setting when generic nonlinear interactions have to be taken into account.

In the following, we concentrate on $U/J=200$ and $\bar{\omega}=\Omega/U = 0.49$, the sign-free red-detuned region near the $\Omega=U/2$ resonance. To assess ground state physics with our finite-temperature method, we employ the standard approach of combined finite-temperature and finite-size scaling $T=J/2L$, so that the temperature scales like the finite-size gap of the system, assuming a dynamical exponent $z=1$ \cite{Matsumoto2001,Ganesh2011}.


\subsection{Results}
\label{sec:photons}
In this section, we present QMC results that shed light on the two main questions of our work: First, does the light-matter coupling change the critical ratio $\JD^c/J$ and shift the position of the QCP? Second, does it change the nature of the QCP itself?

\begin{figure}
	\includegraphics{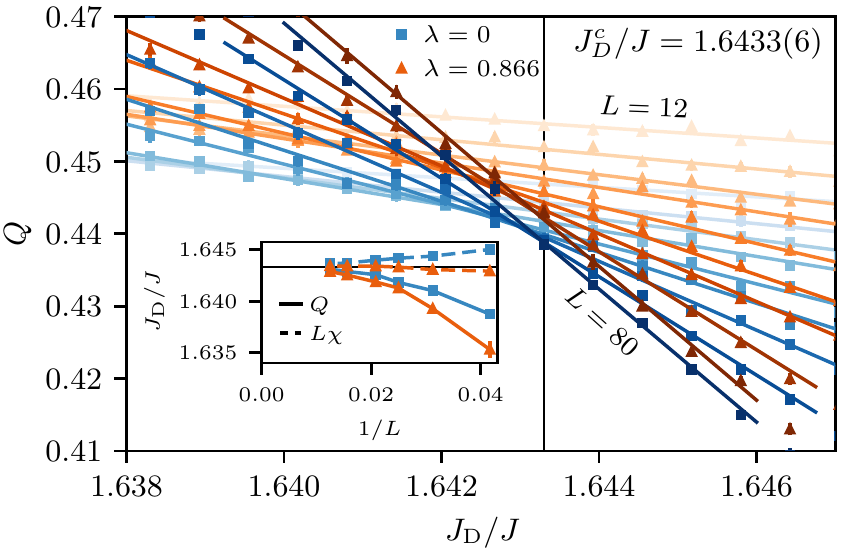}
	\caption{{\bf Numerical determination of the quantum critical point.} The critical coupling ratio $\JD^c/J$ is determined using a finite-size crossing analysis. Shown are two bundles of curves of the Binder ratio $Q$ for zero and finite light-matter coupling, respectively. Each bundle consists of system sizes in the set $L \in \{12, 16, 20, 24, 32, 40, 48, 64, 80\}$. The solid lines show cubic polynomial fits from which crossing points are extracted. The inset shows the crossings of the $Q$ curves at $L$ and $L/2$,  in addition to those of the uniform magnetic susceptibility times system size, $LJ\chi$. From the convergence of the crossings, the position of the critical point $\JD^c/J$ can be determined.}
	\label{fig:honeycomb_crossings}
\end{figure}

To answer the first question we perform a finite-size crossing analysis, i.e. we look at the crossings of observables with known critical finite-size scaling to numerically determine the critical point $\JD^c/J$. A convenient observable for this purpose is the Binder ratio $Q=\braket{m_s^2}^2/\braket{m_s^4}$, where $m_s$ is the staggered magnetization of the AFM order. At the critical point, the scaling of the numerator and the denominator cancel so that the Binder ratio becomes independent of system size. Thus, plotting the Binder ratio for different system sizes $L$ leads to lines crossing at the point where the system displays critical behavior (Fig.~\ref{fig:honeycomb_crossings}).

We extract the crossings between system sizes $L$ and $L/2$ by fitting cubic polynomials to the data. The resulting crossing points still have a small system size dependence due to subleading corrections (inset of Fig.~\ref{fig:honeycomb_crossings}). In addition, the same analysis is carried out for another dimensionless quantity, the uniform magnetic susceptibility multiplied by the system size, $L \chi$. All extracted crossings appear within our resolution to converge to a common limit, $\JD^c/J = 1.6433(6)$, indicating that the $1/\sqrt{N}$ coupling to the cavity mode does not shift the position of the critical point, in agreement earlier scaling arguments~\cite{Andolina2019,Andolina2020,Lenk2022}.

Next, we focus on the QCP itself. Even if the cavity vacuum fluctuations are not strong enough to shift its position, they may still change the nature of the quantum critical ground state in more subtle ways. The nature of a QCP is, in analogy to classical critical phenomena, usually classified by the universal scaling exponents of certain physical observables \cite{sachdev_quantum_2011} that can be extracted from their finite-size scaling.

\begin{figure}
    \centering
    \includegraphics{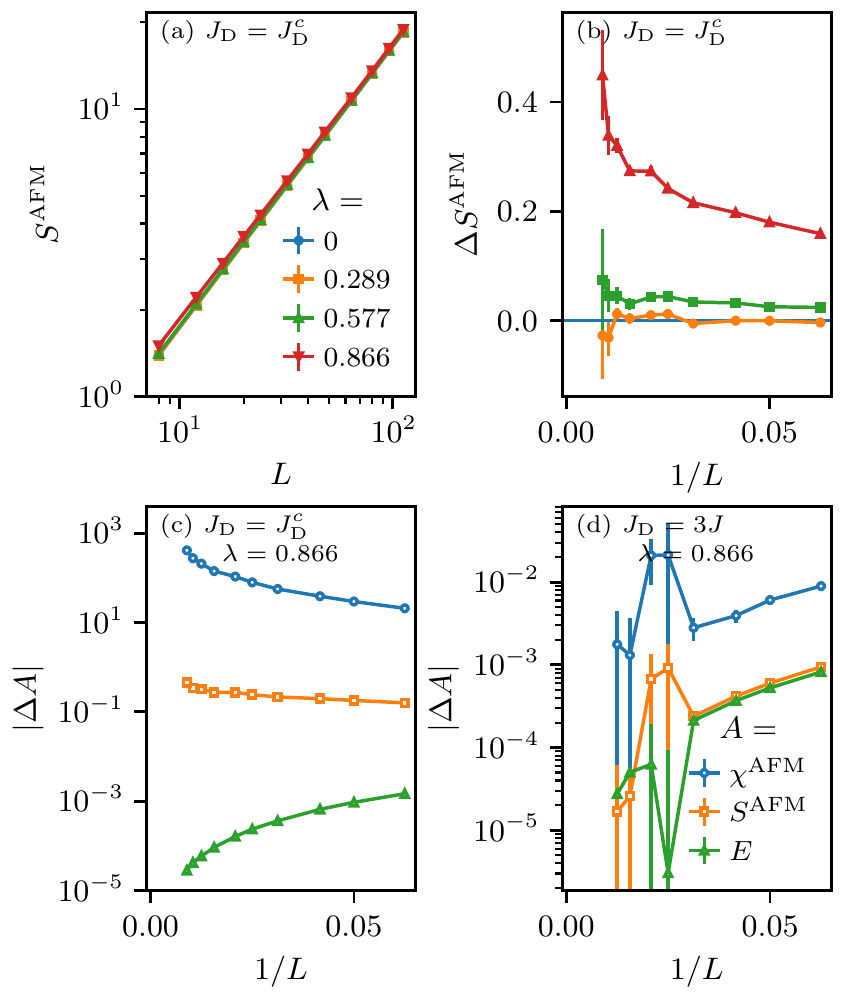}
    \caption{{\bf Cavity effect on the critical scaling of the structure factor and susceptibility.} The absolute enhancement of the antiferromagnetic (AFM) structure factor and susceptibility, $\Safm$ and $\chiafm$, as well as the energy per spin $E$, under the influence of the light-matter coupling $\lambda$ and as a function of system size. (a) The critical scaling of $\Safm$ for different $\lambda$. (b) The absolute difference, $\Delta \Safm = \Safm-\Safm_{\lambda=0}$, for the data in panel (a). (c) Comparison of the absolute difference $\Delta A = A - A_{\lambda=0}$ for different observables $A$ at the critical point. (d) Comparison of the absolute difference for different observables in the dimer-singlet phase.}
    \label{fig:honeycomb_absolute}
\end{figure}

\begin{figure}
	\includegraphics{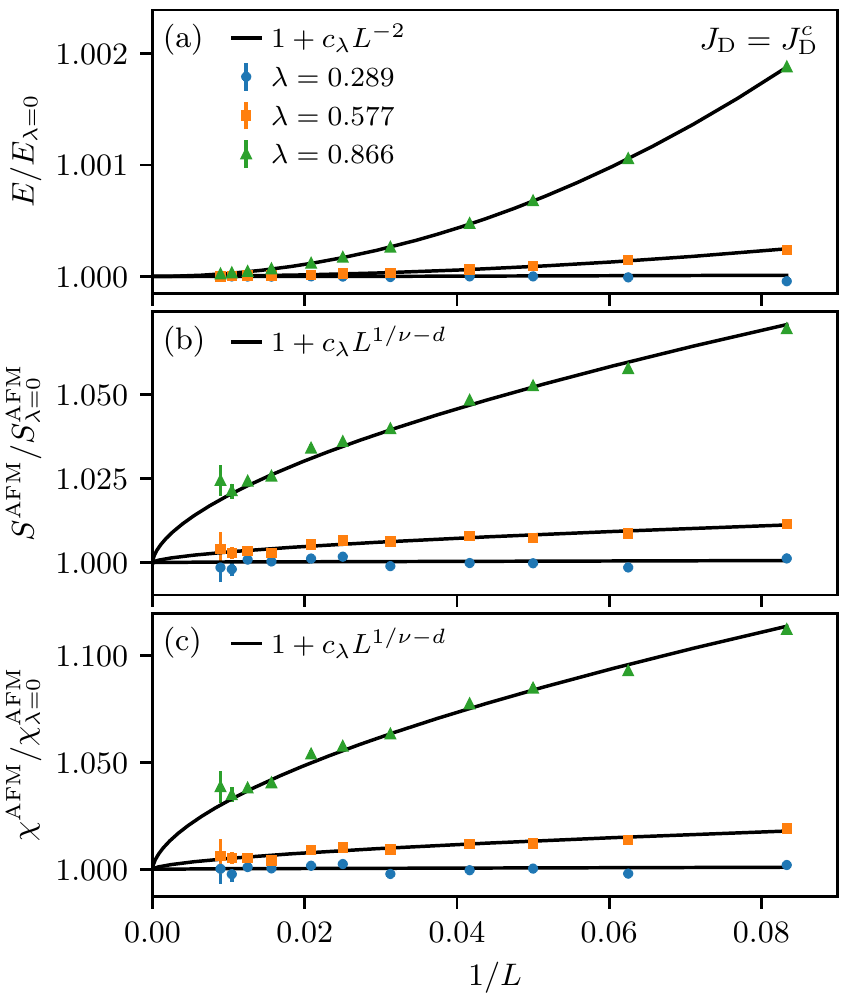}
	\caption{{\bf Cavity effect on the relative critical scaling of the structure factor and susceptibility.} The scaling of different observables normalized by their values for vanishing light-matter coupling, $\lambda=0$. Shown are (a) the energy per spin, $E$, (b) the antiferromagnetic (AFM) structure factor corresponding to the ordering pattern of the transition, and (c) the AFM susceptibility, $\chiafm$. The black lines are fits based on a scaling argument in Sec.~\ref{sec:fields}.}
	\label{fig:scaling_honeycomb}
\end{figure}

To investigate the influence of the cavity on the critical scaling, we calculate the energy per spin $E$ as well as the AFM structure factor and susceptibility
\begin{align}
    \Safm &= \frac{1}{3L^2} \sum_{ij} (-1)^{i+j} \braket{\mathbf{S}_i\cdot \mathbf{S}_j},\\
    \chiafm &= \frac{1}{3L^2} \sum_{ij} (-1)^{i+j} \int_0^\beta d\tau \braket{\mathbf{S}_i(\tau)\cdot \mathbf{S}_j},
\end{align}
where the signs in $\Safm$ are positive/negative on the different magnetic sublattices, and $\tau$ is imaginary time (Fig. \ref{fig:honeycomb_absolute}). The AFM structure factor ($\Safm$) is directly related to the critical fluctuations of the AFM order parameter, and shows an enhancement with increasing coupling to the cavity (Fig. \ref{fig:honeycomb_absolute}(a)). The absolute difference from its $\lambda=0$ value reveals that this enhancement remains in the thermodynamic limit and seems to grow with system size (Fig. \ref{fig:honeycomb_absolute}(b)). A similar picture holds for $\chiafm$, which additionally probes the low-lying excitations above the ground state (Fig.~\ref{fig:honeycomb_absolute}(c)). By contrast, the energy is only weakly enhanced with a vanishing effect for large system sizes. Away from the critical point, the effect of the cavity generally decreases with system size (Fig.~\ref{fig:honeycomb_absolute}(d)).

In part, this behavior is simply due to the different magnitude of the observables itself and due to the fact that $\Safm$ and $\chiafm$ diverge with system size, whereas $E$ converges to a constant. It is therefore instructive to consider the relative enhancement of these quantities as well (Fig.~\ref{fig:scaling_honeycomb}). For the relative enhancement, again, the energy shows the weakest effect, while both $\Safm$ and $\chiafm$ behave qualitatively different from the energy but similar among themselves. In all three cases, the relative enhancements decay in the thermodynamic limit, which means that the single-mode cavity does not change the leading scaling exponents and thus the universality class of the system. Instead it gives rise to what can be interpreted as a “correction to scaling”~\cite{wegner_corrections_1972}, analogously to the corrections to scaling that always appear because of microscopic (or macroscopic~\cite{fradkin_entanglement_2006}) details of the model.

For an effect perturbative in $\lambda$, we would expect such corrections to scale as $\lambda^4/L^2$ as the leading order of our light-matter coupling is $\lambda^2/L^2$ and the effect arises as a back-action from the $\mathcal{O}(\lambda^2)$ cavity vacuum fluctuations onto the matter system. The decay of the energy enhancement fits well with an $L^{-2}$ power-law with $\lambda$-dependent prefactor (Fig.~\ref{fig:scaling_honeycomb}(a)). Choosing the prefactor proportional to $\lambda^4$ is not entirely sufficient to fit the dependence on the light-matter coupling, which we attribute to a small $\mathcal{O}(\lambda^2/N)$ direct renormalization of the exchange coupling in the $\braket{0|\Jph|0}$ matrix element (see App.~\ref{app:bond_op}).

However, in the presence of the singular behavior at a QCP, such simple perturbative arguments need not always hold true. This is illustrated here in the case of the magnetic fluctuations. Here, the power-law decay is better described by a much smaller exponent, compatible with $1/\nu-d = -0.596(5)$ (based on $\nu=0.7121(28)$ for the 3D O(3) class~\cite{Kos2016}), which we derive based on a scaling argument in the Sec.~\ref{sec:fields}. In App.~\ref{app:cd_lattice}, we show that the same exponent also appears in a different lattice featuring an AFM-dimer-singlet QCP.

At larger system sizes, this behavior may be modified due to the relevance of additional finite-$q$ cavity modes. In App.~\ref{app:multimode}, we show within the field theory description to be discussed below that the inclusion of multiple modes leads to a finite shift of the QCP and a stronger non-local interaction term.


\subsection{State of the cavity}
\begin{figure}
	\includegraphics{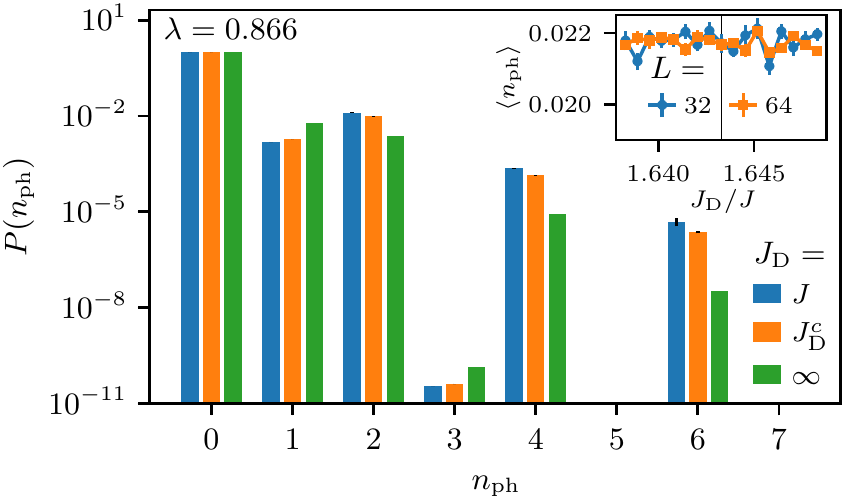}
	\caption{{\bf Cavity mode occupation of the coupled spin-photon system.} Occupation number distribution $P(n_\text{ph})$ of the cavity photon at different values of $\JD$ including the observable correction arising from the downfolding. The results for $L=32$ shown here are converged to the thermodynamic limit as is visible in the inset, showing the average occupation $\braket{n_\text{ph}}$ as a function of $\JD/J$ around the critical point (vertical line).}
	\label{fig:photon_dist}
\end{figure}

So far, we have discussed observables of the matter system. The state of the cavity is also accessible to our QMC formulation via the occupation number distribution, $P(n_\text{ph}) = \braket{\ket{n_\text{ph}}\!\bra{n_\text{ph}}}$, (Fig.~\ref{fig:photon_dist}). Due to the downfolding, the photon states in our model do not exactly correspond to the physical photons, so that $P(n_\text{ph}) = P_\text{eff}(n_\text{ph}) + \Delta P(n_\text{ph})$ is subject to a small correction term that we derive in App.~\ref{app:ed_comparison}).

In addition to numerical results for $\JD=J$ and $\JD=\JD^c$, we include analytical results obtained for the case $\JD=\infty$ where the spin state of the model becomes an exact singlet product. 

Two contributions on the occupation number distribution can be separated. First, the light-matter coupling within our effective model leads a virtual occupation of the even-numbered photon sector. Second, an overall smaller contribution enters for all $n_\text{ph}$ due to the correction $\Delta P(n_\text{ph})$, dominating in the odd-numbered sector. Considering the parity symmetry, these odd-numbered states are likely a sign of light-matter entanglement similar to the ones recently found in a one-dimensional interacting model~\cite{Passetti2022}.
We find that the static occupation number distribution does not show a distinct signature at the critical point (inset of Fig.~\ref{fig:photon_dist}).

In principle, like for an impurity in a bulk system, the critical magnet should mediate long-time correlations that could be used as a cavity probe for critical behavior. Such correlations are, as we have shown, not visible in the static observables easily accessible in our QMC simulations. We do, however, expect them to appear in dynamical observables such as the second-order degree of coherence $g^{(2)}(t)$.


\section{Field-theory picture}
\label{sec:fields}
In the preceding section, we presented numerical results for the light-matter enhancement of different observables, which had several key properties: (i) The enhancement is strong for magnetic observables. (ii) The relative enhancement, when viewed as a correction to scaling, has an exponent that is similar for different observables and lattices.
These properties suggest that the enhancement effect can be understood analytically through the lense of a field theory that has a more universal scope than our particular microscopic model.

The starting point of this idea is to perform the continuum limit of our lattice model, which is done in the framework of bond operator theory~\cite{Sachdev1990,Kotov1998} in App.~\ref{app:bond_op}. For this limit, we assume from the start that the photon occupation is always low so that the higher-order terms of $\Jph$ can be dropped. Furthermore, we drop higher-order magnetic interactions which are irrelevant (in the renormalization group sense) close to the critical point. These considerations lead to the action
\begin{align}
	\label{eq:generic_action}
	\mathcal{S}_{\rm s-ph} &= \int d\tau \Big (a^*\partial_\tau a + \Omega |a|^2 + \lambda^2 \Gamma_0 \Rea a^2\nonumber\\
	 &+ \int d^dx \Big[ -\frac{1}{2}\bphi\cdot (\nabla^2 - g)\bphi + \frac{u}{4} (\bphi^2)^2 \nonumber\\
	 &+ \frac{\lambda^2}{L^d} \left(\Gamma_1 |a|^2 + \Gamma_2 \Rea a^2 + \Gamma_3\right)\bphi^2 \Big] \Big),
\end{align}
where $d$ is the spatial dimension, $\tau$ is imaginary time, $a$ is a complex field describing the cavity photon, and $\bphi$ is a real vector field describing the coarse-grained AFM order parameter. In addition to the terms presented here, $\Omega$ is shifted by a term of $\mathcal{O}(\lambda^2)$, which does not affect our results at leading order in $\lambda^2$. While derived from our specific microscopic model in Eq.~\ref{eq:spinham}, we stress that this action is quite generic. It could in fact, based on symmetry considerations alone, have been written down phenomenologically for any O($N$) critical point coupled quadratically to a single bosonic mode.

To understand the effect of the photon mode, it can be integrated out to leading order in $\lambda$ (see App.~\ref{app:bond_op}), yielding the standard O($N$) model
\begin{align}
	\label{eq:onmodel}
	\mathcal{S}_{\rm s} &=\int d\tau d^dx \Bigg[ -\frac{1}{2}\bphi\cdot \left(\nabla^2 - g + \frac{\Sigma}{L^d}\right) \bphi\nonumber\\&+ \bphi^2 \left(\frac{u}{4} \bphi^2 + \int d\tau' d^dy \frac{V(\tau-\tau')}{L^{2d}} \bphi^2(\tau', y)\right)\Bigg],
\end{align}
where both the mass and the interaction terms acquire corrections, $\Sigma=\lambda^2\Gamma_3+\lambda^4 \Gamma_0\Gamma_2/\Omega$ and $V = -2\lambda^4 \Gamma_2^2/\Omega$. Intuitively, since at the critical point the mass term in the original (uncoupled) model is zero, $\Sigma$ can have a strong effect even though it is suppressed by a volume factor $1/L^d$. The modified interaction $V$, on the other hand, is always a small addition on top of the existing quartic interactions, although its non-local nature could have consequences as well.

In the following, let us consider an observable $A$ close to the critical point. If $A$ is singular (as e.g. the structure factor), it will assume a power-law form $A \sim g^p$ with some observable dependent exponent $p$. For $\lambda \ne 0$,  $g$ in this form is replaced by $g + \Sigma L^{-d}$. Further, $g$ is related to the correlation length, which in turn is cut off by $L$ for a finite system at the critical point, so that $g\sim L^{-1/\nu}$. Then for any exponent $p$ we get
\begin{align}
	\label{eq:scaling_ratio}
	\frac{A}{A_{\lambda=0}} \sim \frac{\left(L^{-1/\nu} + \Sigma L^{-d}\right)^p}{L^{-p/\nu}} \sim 1 + c' L^{1/\nu-d}.
\end{align}
For $d=2$ and $\nu = 0.7121(28)$ \cite{Kos2016} in the $(2+1)$D O(3) universality class, the value of this correction exponent is $1/\nu-d = -0.596(5)$ which fits well with our data (Fig.~\ref{fig:scaling_honeycomb}), while also explaining the similarity of the correction exponents for different observables. The absolute difference $A-A_{\lambda=0} \sim L^{(1-p)/\nu -d}$ diverges or vanishes depending on the observable. In particular, for $\Safm$, $p=2\beta-\nu d$ and $(1 - p)\nu - d \approx 0.366(6) > 0$.

The appearance of these exponents is actually quite unexpected and special to the strongly correlated nature of the system. In most situations, one would expect that a perturbative expansion in the light matter coupling, $A \approx A^{(0)} + A^{(1)} \Sigma L^{-d}$ exists so that the light-matter enhancement scales like $L^{-d}$. Such considerations form the basis of many arguments about the strength of a single mode in weakly correlated systems. Tuning the matter system to a QCP makes the $\lambda\to 0$ limit singular, breaking simple perturbative arguments and giving rise to a stronger than expected non-analytic scaling.
Finally, for observables either (i) dominated by their non-singular part such as the energy or (ii) far away from the critical point, the simple perturbative expansion works again and the $L^{-d}$ scaling is recovered. 

The exponent $1/\nu-d$ of the relative enhancement further suggests that the effect is stronger in other universality classes. For example, in the (1+1)D Ising model, $d=\nu=1$ leading to a constant effect in the thermodynamic limit. For the (1+1)D three-state Potts universality class, $\nu = 5/6 < 1$~\cite{Pearson1980}, so that the correction diverges with system size, signalling a shift of the critical coupling or a change in the leading critical exponents.


\section{Conclusion}
\label{sec:conclusion}
We have studied a quantum critical magnet coupled to a single-mode cavity in the dipole approximation using large-scale QMC simulations. Our results show that while the position and universality class of the quantum critical point are not changed, the single mode has an influence on observables related to the critical magnetic fluctuations in the magnet.
Using a scaling argument, this influence can be viewed as a correction to the critical scaling with an exponent $1/\nu-d$ that is independent on the microscopic details of the lattice. As a result, in our case, the relative enhancement of the fluctuations tends to zero in the thermodynamic limit. For certain observables such as the static AFM structure factor, the absolute enhancement, however, still diverges in the thermodynamic limit, which can be seen as a change in the ground state of the matter system, induced by a single cavity mode. On a fundamental level, the emergence of a fractional scaling exponent in the light-matter enhancement highlights that strong correlations coupled to light can induce qualitatively different behavior that falls beyond simple perturbative arguments applicable in weakly correlated systems.

A possible platform to realize our findings experimentally is the van der Waals magnet \mnp{Se}, where the N\'eel AFM to dimer transition could be realized by applying hydrostatic pressure. Here, the renormalization of the AFM order parameter should be accessible by optical probes such as linear dichroism, reflectivity anisotropy, or Raman measurements. Further, while we show that the static photon number statistics are not sensitive to critical fluctuations, we expect the dynamical photon correlations to show a signal of the critical slowing down at the QCP.

In the true thermodynamic limit, additional finite-momentum modes should be taken into account. Including more modes in the QMC method, coupled via the downfolded Peierls coupling, comes with two challenges. First, it generally introduces a sign problem. Second, due to the structure of the coupling, all modes are coupled to each other in a complicated way that leads to an exponential scaling in memory for the current SSE formulation. Both problems may be tackled by a change of the computational basis or further controlled approximations of the coupling. However, including finite-momentum modes in the microscopically derived field theory is straight-forward and leads to a finite shift of the QCP.

Lastly, we stress that the field-theory picture of the physics here is quite robust to the microscopic details of the model and should apply also in other critical systems. While the exponent $1/\nu-d$ is negative in the (2+1)D O(3) universality class we considered, this is not true for other phase transitions. In the (1+1)D Ising class, $\nu=1$~\cite{Kaufman1949} so that the exponent is exactly zero, leading to a constant relative enhancement of observables. In the (1+1)D Potts universality class, $\nu=5/6$~\cite{Pearson1980} so that the exponent becomes positive and dominant over the original critical behavior.

\begin{acknowledgments}
The authors thank Christian Eckhardt, Simone Latini, Marios Michael, and Frank Schlawin for helpful discussions. L. W. acknowledges support by the Deutsche Forschungsgemeinschaft (DFG, German Research Foundation) through grant WE 7176-1-1. This work was supported by the European Research Council (ERC-2015-AdG694097), the Cluster of Excellence ‘Advanced Imaging of Matter' (AIM), Grupos Consolidados (IT1453-22) and Deutsche Forschungsgemeinschaft (DFG) – SFB-925 – project 170620586. We acknowledge support by the Deutsche Forschungsgemeinschaft (DFG, German Research Foundation) via  Germany’s Excellence Strategy -- Cluster of Excellence Matter and Light for Quantum Computing (ML4Q) EXC 2004/1 -- 390534769 and within the RTG 1995. We acknowledge support by the Max Planck-New York City Center for Nonequilibrium Quantum Phenomena. The Flatiron Institute is a division of the Simons Foundation. 
\end{acknowledgments}
\appendix
\section{Comparison to exact diagonalization of the Hubbard model}
\label{app:ed_comparison}
In this appendix, we will consider the $4\times 4$ square lattice as a benchmark system where both simulations of original Hubbard model with Peierls substitution, Eq.~\eqref{eq:hubbard}, and the downfolded Heisenberg model, Eq.~\eqref{eq:spinham}, are feasible. While it should be stressed that the physics of this system is very different from the large-scale, quantum critical systems studied in the remainder of this work, it nevertheless allows us to shed light on some intricacies in the Schrieffer-Wolff transformation that is used to get to the effective spin description.

\begin{figure}
    \centering
    \includegraphics{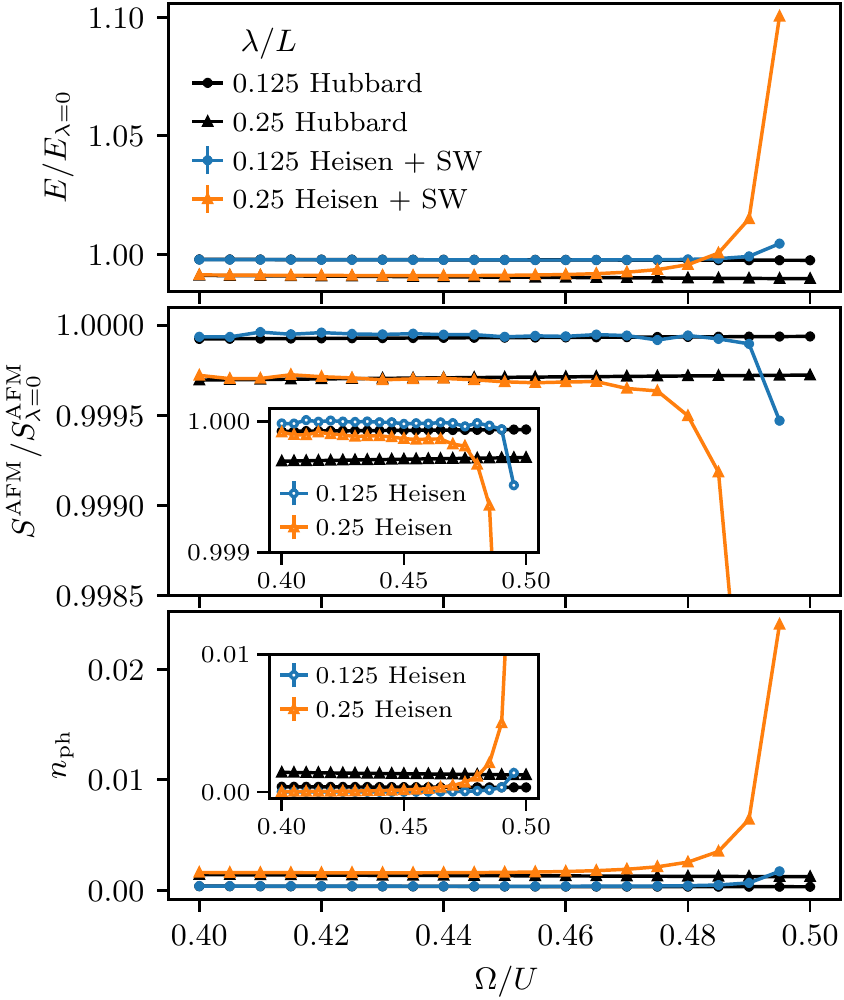}
    \caption{Comparison of the light-induced change of the energy $E$, the antiferromagnetic structure factor $\Safm$, and the photon occupation number $n_\text{ph}$. The main panels show a comparison of ED results for the Hubbard model and QMC results for the downfolded Heisenberg model including a correction from the Schrieffer-Wolff transformation (Heisen + SW). Both model are realized on the square lattice with $U/J=200$. The insets show the Heisenberg results without the correction (Heisen).}
    \label{fig:ed_comparison}
\end{figure}
The Hubbard model results were obtained using ED for the ground state, while the Heisenberg model was simulated using the same QMC method as in the main text at a temperature of $T/J=0.05$, which is sufficient to converge to the ground state. Fig.~\ref{fig:ed_comparison} results for the Energy, AFM structure factor and photon occupation in the two models.

The results for the downfolded model show two kinds of discrepancies from the Hubbard model. First, close to $\Omega/U = 1/2$, the downfolded observables diverge. This is due to the singularity appearing in the downfolded exchange coupling signaling a breakdown of the underlying perturbation theory. However, the spread of this divergence decreases with $\lambda/L$. We therefore expect that this issue will be less severe at $\Omega/U = 0.49$ and the larger system sizes we studied. In future works, comparing the Hubbard model and the downfolded Heisenberg model in this regime and finding an improved approximation without the singularities would be of interest.

The second kind of discrepancy between the downfolded Heisenberg and the Hubbard model can be seen for all $\Omega/U$ in the insets of Fig.~\ref{fig:ed_comparison} and stems from the fact that the observables in the downfolded model are not exactly equivalent to those in the Hubbard model but, like the effective Hamiltonian, differ by a Schrieffer-Wolff transform,
\begin{align}
    H_\text{eff} &= e^{S} H e^{-S}\\
    H_\text{eff}\ket{\Phi} &= E \ket{\Phi}\\
    \Rightarrow H \underbrace{e^{-S} \ket{\Phi}}_{=\ket{\Psi}} &= E \underbrace{e^{-S}\ket{\Phi}}_{=\ket{\Psi}}\\
    \Rightarrow \braket{\Psi| A |\Psi} &= \braket{\Phi|e^{S} A e^{-S}|\Phi},
\end{align}
where $S$ is the antihermitian operator of the Schrieffer-Wolff transform that is chosen to perturbatively eliminate the offdiagonal part of $H$.

In many common cases, such as magnetic observables in the Heisenberg model, any contribution coming from $e^{S}$ will be at higher order in perturbation theory, so that $\braket{\Phi|A|\Phi} \approx \braket{\Psi|A|\Psi}$. For example, the leading order of $\Safm$ is $(t/U)^0$. However, if we consider light-induced enhancements, we often encounter effects of order $J/\Omega = t^2/U\Omega$, which are of a magnitude similar to the leading contribution $t^2/U^2$ from the transformation. These corrections can be calculated starting from the expression for $S$ in our case,
\begin{align}
S_{m n; l n'} = - \sum_{\braket{ij}} t_{ij} \frac{\braket{m|c_i^\dagger c_j D^{ij}_{nn'} + \text{h.c.} |l}}{U + \Omega (n'-n)},
\end{align}
where $m$ enumerates states in the zero-doublon subspace and $l$ enumerates states in rest of the Hilbert space. $n$, $n'$ are arbitrary photon numbers. Antihermiticity determines the rest of the matrix elements of $S$.

To compare our results to ED, we compute in the following the correction $\braket{\Delta A} = \braket{\Psi|A|\Psi} - \braket{\Phi|A|\Phi}$ for diagonal photonic observables like $n_\text{ph}$ and for the AFM structure factor $\Safm$ to leading in perturbation theory. This leading order, due to the inclusion of higher order of $S$ can be computed using the leading order of $\ket{\Phi}$, which has zero photon occupation $\ket{\Phi} = \ket{\phi}\otimes \ket{n_\text{ph}=0}$.

The correction for the photon number $n_\text{ph}$ or in general, an observable $f(n_\text{ph})$ with $f(0)=0$ then becomes
\begin{align}
\Delta f(n_\text{ph})= & -\sum_{\braket{ij}}  \underbrace{\sum_{n=0}^\infty\frac{f(n) |D^{ij}_{0n}|^2}{(1+\omega n)^2}}_{=: R_{ij}[f]} \frac{J_{ij}}{U} \left(\mathbf{S}_i \cdot \mathbf{S}_j - \frac 14\right).
\end{align}
We do not have direct access to this operator, so we will approximate it in terms of known quantities,
\begin{align}
\label{eq:photon_correction}
\braket{\Delta f(n_\text{ph})}\approx -\frac{E}{U} \sum_{\braket{ij}} \frac{J_{ij}}{\sum_{\braket{ij}} J_{ij}} R_{ij}[f].
\end{align}
This approximation essentially assumes that the energy is equally distributed on the bond terms in the Hamiltonian so that bond expectation values can be written as a weighted fraction of the energy.

Eq.~\eqref{eq:photon_correction} can also be used to calculate the correction of the occupation number distribution $P(n_\text{ph})$ for $n_\text{ph}>0$. By normalization, $\Delta P(n_\text{ph}=0) = -\sum_{n_{\text{ph}}>0} \Delta P(n_\text{ph})$.

For $\Safm$ we consider the correction to the light-induced difference only.
\begin{align}
&\Delta \Safm - \Delta \Safm_{\lambda=0} =\nonumber\\
&\frac{1}{2} \left\{\Safm, \sum_{\braket{ij}} R_{ij}[g] \frac{J_{ij}}{U} \left(\mathbf{S}_i \cdot \mathbf{S}_j -\frac 1 4\right)\right\}\nonumber\\
&+ \sum_{\braket{ij}} \frac{t_{ij}^2}{U^2} \sum_{n=1}^\infty P \frac{c_i^\dagger c_j D^{ij}_{0n}+\text{h.c.}}{1+\omega n} \Safm Q \frac{c_i^\dagger c_j D^{ij}_{0n}+\text{h.c.}}{1+\omega n} P.
\end{align}
Here, $g(n) = 1$ if $n>0$ and $0$ for $n=0$. $P$ is a projector on the zero-doublon subspace and $Q$ is a projector on the one-doublon subspace. This operator is complicated to evaluate, but we can find a crude approximation by assuming that $\Safm$ is almost the same in the one-doublon subspace, up to two missing zero-spin sites (the holon and the doublon) that have been deleted by the fermionic operators. Apart from this we neglect all correlations between $\Safm$ and other operators, so that we can apply again a similar approximation as for the photonic observables.
\begin{align}
&\braket{\Delta \Safm - \Delta \Safm_{\lambda=0}}\nonumber\\
&\approx\frac 12 \Braket{\left\{\Safm, B\right\}} - \frac{N-2}{N} \braket{\Safm} \braket{B}\nonumber\\
&\approx\frac{2}{N} \braket{\Safm} \braket{B}\nonumber\\
&\approx\frac{2 E \braket{\Safm}}{NU} \sum_{\braket{ij}} \frac{J_{ij}}{\sum_{\braket{ij}} J_{ij}} R_{ij}[g],
\end{align}
with $B= \sum_{\braket{ij}} R_{ij}[g] J_{ij} (\mathbf{S}_i\cdot\mathbf{S}_j - 1/4)/U$.

Despite these approximations, adding these correction terms leads to a remarkable agreement with the ED results (Fig.~\ref{fig:ed_comparison}).

We find that the corrections, while dominant in the $4\times 4$ square lattice, are small compared to the bare observables $\braket{\Phi|A|\Phi}$ in the larger lattices, so we drop them in the other calculations.

\section{Derivation of the continuum action in dimerized antiferromagnets}
\label{app:bond_op}
To elucidate our numerical findings further, in this section, we will develop a complementary analytical approach based on a field theoretical scaling argument. Starting from the specific microscopic Hamiltonian~\eqref{eq:spinham}, we derive a continuum action describing the physics of the magnetic quantum critical point coupled to a cavity photon. Upon integrating out the photon, we recover the well-known O(N) model with the addition of a cavity-induced mass term, which generally vanishes with increasing system sizes but becomes relevant close to the critical point.

The influence of the cavity-induced mass term will allow us to explain the scaling of the cavity-induced enhancements observed in our numerical results of Sec.~\ref{sec:qmc}. While this approach starts from our specific model, the universal nature of the field-theory description allows us to anticipate under what conditions the same physics may be found also in other systems hosting a O($N$) QCP.

The basic assumptions underlying this analysis is, first, that our model is close enough to the critical point to be described by a continuum theory of the lowest-energy excitations. Second, we assume (as we confirm in Sec.~\ref{sec:photons}) that the photon occupation is low so that we can neglect higher order terms in the light-matter coupling.

To start, we consider the physics of our AFM close to the QCP separating the dimer-singlet and AFM phases. At this QCP, the relevant critical fluctuations are triplet excitations that break the dimer singlets of the disordered phase and condense to form AFM order. To describe these excitations in a bosonic language, we use the bond-operator formalism~\cite{Sachdev1990,Kotov1998}, which can be considered a version of spin-wave theory for dimer-singlet ground states.

The eigenbasis of a single dimer, consisting one singlet and three triplet states, can be written as
{
	\def\u{\uparrow}
	\def\d{\downarrow}
\begin{align}
\ket{s} &= \phantom{i}(\ket{\u\d}-\ket{\d\u})/\sqrt{2},\nonumber\\
\ket{t_x} &= \phantom{i}(\ket{\d\d}-\ket{\u\u})/\sqrt{2},\nonumber\\
\ket{t_y} &= i(\ket{\d\d}+\ket{\u\u})/\sqrt{2},\nonumber\\
\ket{t_z} &= \phantom{i}(\ket{\u\d}+\ket{\d\u})/\sqrt{2}.
\end{align}
}
For these states, three bosonic bond operators are defined that create triplet states from the singlet “vacuum”
\begin{equation}
	\td_a\ket{s} = \ket{t_a},\quad a=x,y,z
\end{equation}
and fulfill bosonic commutation relations.

To avoid unphysical states, the new bosonic Hilbert space has then to be constrained to the sector $\btd\cdot\bt \le 1$. Where $\bt$ is the vector of $t_a$ operators, transforming like a vector under spin rotations.

In this language, the two spin operators belonging to the dimer can be expressed as
\begin{equation}
	\label{eq:spindec}
\mathbf{S}_{1,2} = \frac 12 (\pm \btd P \pm P \bt - i \btd\times\bt),
\end{equation}
where $P=1-\btd\cdot\bt$ is a projector onto the physical subspace.

The dimerized magnet we are interested in is made up of $\JD$ intradimer and $J$ interdimer bonds. For each $\JD$ bond $b$ we introduce a set of bond operators $\bt_b$. Then, we express the Hamiltonian containing all bonds using Eq.~\eqref{eq:spindec}. In a bilayer geometry~\cite{Joshi2015}, one gets $H = H_0 + H_2 + H_4 + \mathcal{O}(\bt^6)$ with
\begin{align}
	H_0 &= \Omega a^\dagger a -\frac{3N\JD}{4} - \sum_{\braket{bd}} \frac{\Jph_{bd}}{4},\nonumber\\
	H_2 &= \JD\sum_b \btd_b \cdot \bt_b + \sum_{\braket{bd}} \frac{\Jph_{bd}}{2} \normord{(\btd_b + \bt_b) \cdot(\btd_d + \bt_d)}\nonumber\\
	H_4 &= \sum_{\braket{bd}} \frac{\Jph_{bd}}{2}\Big(\normord{(\btd_{b}\cdot\bt_{d})  (\btd_{d}\cdot\bt_{b})} - (\btd_b\cdot\btd_d)(\bt_b\cdot\bt_d))\nonumber\\
	&- \normord{\left[(\btd_{b} + \bt_{b})\cdot(\btd_{d} + \bt_{d})\right] (\btd_{b}\cdot \bt_{b} + \btd_{d}\cdot\bt_d)}
	 \Big)
\end{align}
where the sums count neighboring bonds and some terms contain normal ordering for brevity. The expansion also produces sixth-order terms in $\bt$, but we shall ignore them in the following analysis, where $\braket{\btd\cdot\bt}$ is always assumed to be small.
In a similar way, we drop terms $\mathcal{O}((\Jph-J)\bt^4)$ that are suppressed by low cavity occupation.

\begin{align}
	Z = \int\mathcal{D}\bt_b^*\mathcal{D}\bt_b \mathcal{D}a^* \mathcal{D}a\,e^{-\mathcal{S}[\bt_b^*,\bt_b, a^*, a]}
\end{align}
with $\mathcal{S} = \int_0^\beta d\tau\, a^*\partial_\tau a + \sum_b \bt_b^*\cdot \partial_\tau \bt_b + H[\bt_b^*,\bt_b, a^*, a] = \mathcal{S}_0 + \mathcal{S}_2 + \mathcal{S}_4 $.

Due to the bipartiteness of our bilayer, the purely bilinear part of $\mathcal{S}_2$ is minimized by configurations following the sign structure of the AFM, $\bt_b \propto (-1)^b$, where $(-1)^b$ is negative on one sublattice and positive on the other. The low-energy theory including this mode and its low-energy excitations can be obtained by performing the continuum limit $\bt_b \approx (-1)^b \bt(r)$, where $\bt(r)$ is a slowly varying function that can be expanded to second order in the bond length.
Ignoring derivatives in both matter-matter and light-matter interaction terms, this leads to the action
\begin{align}
	\mathcal{S}_0 &= \int d\tau\, a^*\partial_\tau a + \Omega |a|^2 - \frac{zN}{8} \Jph(a^*,a)+ \text{const},\nonumber\\ 
	\mathcal{S}_2 &= \int d\tau d^d x\,\bt^* \cdot\partial_\tau \bt\nonumber\\
	&\qquad- (\bt^* + \bt) \cdot \nabla^2 (\bt^* + \bt)\nonumber\\
	&\qquad+ B_1\left[\JD |\bt|^2 - \frac{z\Jph(a^*,a)}{4} (\bt^* + \bt)^2\right],\nonumber\\
	\mathcal{S}_4 &= \int d\tau d^d x\,B_2 \begin{aligned}[t]
		\big[&(|\bt|^2)^2-|\bt^2|^2\\
	- &2 (\bt^* + \bt)^2 |\bt|^2\big].
	\end{aligned}
\end{align}
Here, the isotropic form of the derivative term has been fixed by performing transformation of the coordinates and fields. This gives rise to the lattice dependent constants $B_1$ and $B_2$. The coupling $\Jph$ arises from bond-averaging the lattice-dependent coupling,
\begin{align}
	\Jph(a^*,a) &= \frac{1}{z} \sum_{\braket{0,d}} \Jph_{0d}(a^*,a),
\end{align}
where $z$ is the number of nearest neighbor dimers.

\begin{figure}
    \centering
    \includegraphics{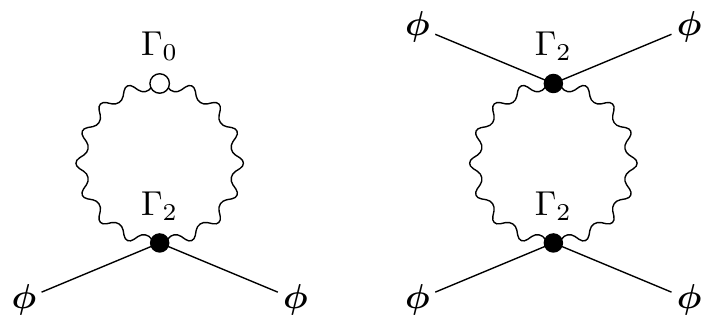}
    \caption{The diagrams that contribute when integrating out the photons at $T=0$ up to order $\lambda^4$.}
    \label{fig:diagrams}
\end{figure}

Next, we express the the complex field $\bt$ in terms of real fields $\bt = \bphi +  i \bpi$, and noting that $\bpi$ is always gapped, we integrate it out, obtaining
\begin{align}
	\mathcal{S} &= \int d\tau a^*\partial_\tau a + \Omega |a|^2 - \frac{zN}{8}\Jph\nonumber\\&+\int d\tau d^d x\, - \frac{1}{2} \bphi \cdot (\partial_\tau^2 + \nabla^2 + g) \bphi\nonumber\\
	&\phantom{+\int d\tau d^d x} + B_1 z(\Jph-J) \bphi^2 + \frac{u}{4} (\bphi^2)^2.
\end{align}
After approximating $\Jph$ to quadratic order in $\lambda$, we arrive at the action from Eq.~\eqref{eq:generic_action} in the main text with 
\begin{align}
    \Gamma_0&= -8 B_3 \Gamma_2,\nonumber\\
    \Gamma_1&= \alpha \left( \frac{4}{1-\bar{\omega}} - \frac{2}{1-4\bar{\omega}^2} - 2\right),\nonumber\\
    \Gamma_2 &= \alpha \left(\frac{2}{1-\bar{\omega}} + \frac{1}{1-4\bar{\omega}^2} - 1\right),\nonumber\\
    \Gamma_3 &= \alpha\left(\frac{2}{1+\bar{\omega}}-1\right),
\end{align}
where $\alpha = B_1 J \sum_{\braket{0,d}} \hat{\mathbf{r}}_{0,d}\cdot \boldsymbol{\epsilon}$ contains a sum over the polarization projected on the bond directions. $B_3$ is another geometrical factor.
From there, the photons can be integrated out perturbatively up to order $\lambda^4/L^d$. At this order and $T=0$, only processes creating virtual photons (i.e. those not containing $\Gamma_1$) contribute (Fig.~\ref{fig:diagrams}). This results in the O($N$) model of Eq.~\eqref{eq:onmodel}.
\section{Multiple cavity modes}
\label{app:multimode}
In the main text, we consider only a single mode in dipole approximation, which allows us to solve the microscopic model using QMC simulations. This solution, we rationalize using a scaling argument based on a continuum field theory. In this appendix, we take a complementary approach: In the continuum field theory, whose applicability we have checked for a single mode, we can readily include more cavity modes. The additional modes then eliminate the $1/N$ factors from the light-matter coupling and the effect of the cavity on the matter is expected to be much more substantial in the thermodynamic limit.

The derivation of the continuum model follows the same steps, apart from two substitutions in the microscopic Hamiltonian. The photon energy
$\Omega a^\dag a$ should be replaced by a sum $\Omega \vec{a}^\dag \cdot \vec{a}$ and the full downfolded Peierls coupling becomes
\begin{align}
    \braket{\vec{n}|\Jph_{ij}|\vec{m}} &= \frac{J_{ij}}{2} \sum_{\vec{l}} \Rea D^{ij}_{\vec{n}\vec{l}}(D^{ij}_{\vec{l}\vec{m}})^* \nonumber\\
    &\times \left(\frac{1}{1+\vec{\bar{\omega}}\cdot (\vec{l}-\vec{n})} + (\vec{n}\leftrightarrow \vec{m})\right)
\end{align}
where the vectors contain occupation numbers and frequencies for the different modes. The displacement operator,
\begin{align}
    D_{\vec{n}\vec{m}}^{ij} = \braket{\vec{n}|\exp\left[\vec{\lambda}^*\cdot \vec{a}^\dag +  \vec{\lambda}\cdot \vec{a}\right]|\vec{m}},
\end{align}
similarly is replaced by the multiple mode equivalent.

In the following, we will assume a cavity that does not break translation symmetry in the transversal direction and include a set of low-energy modes with small but finite in-plane momentum $|q| < \Lambda$, much smaller than the inverse lattice spacing. The light-matter coupling can then be written as
\begin{align}
    \lambda_{q\zeta}^{ij} \approx \lambda_q (r_i-r_j) \cdot \epsilon_{q\zeta} e^{i q\cdot r_{i}}
\end{align}
with the normalized polarization vectors $\epsilon_{q\zeta}$.
Expanding the downfolded Peierls coupling to second order in $\lambda$ yields
\begin{align}
    \frac{\Jph_{ij}}{J}-1 \approx \frac{1}{2N} \bigg(\frac{1}{2}&\vec{\alpha}^\dag \cdot A_2 \vec{\alpha}^\dag + \frac 1 2 \vec{\alpha} \cdot A_2 \vec{\alpha}\nonumber\\
    +\,&\vec{\alpha}^\dag \cdot A_1 \vec{\alpha}  + \vec{\lambda}^{ij*}\cdot A_3 \vec{\lambda}^{ij}\bigg).
\end{align}
where $\alpha_{q\zeta} = \lambda^{ij}_{q\zeta} a_{q\zeta}$ and
\begin{align}
    A_1^{q\zeta,q'\zeta'} &= \frac{2}{1-\bar{\omega}_q^2} + \frac{2}{1-\bar{\omega}_{q'}^2} - \frac{2}{1-(\bar{\omega}_{q}-\bar{\omega}_{q'})^2} - 2,\nonumber\\
    A_2^{q\zeta,q'\zeta'} &= \frac{1}{1-\bar{\omega}_q} + \frac{1}{1-\bar{\omega}_{q'}} + \frac{1}{1-(\bar{\omega}_q+\bar{\omega}_{q'})^2} - 1,\nonumber\\
    A_3^{q\zeta,q'\zeta'}&= \delta_{qq'} \delta_{\zeta \zeta'} \left(\frac{2}{1+\bar{\omega}_q}-1\right).
\end{align}
Like in the single mode case, we need to perform the bond-direction average, where we observe that for highly symmetric lattices like the honeycomb or square lattice, the product of coupling constants can be simplified to
\begin{align}
    \sum_{j \in\text{n.n.}} \lambda^{ij}_{q\zeta} \lambda^{ij}_{q'\zeta'} = \eta \lambda_{q} \lambda_{q'}\,\epsilon_{q\zeta} \cdot P_\parallel \epsilon_{q'\zeta'}\, e^{i(q+q')\cdot r_i}
\end{align}
where geometric proportionality factor  $\eta$ appears. $P_\parallel$ is a projector on the in-plane component of the polarization, for which we assume $\epsilon_q \cdot P_\parallel \epsilon_{q'}\approx 1$ since we consider only small transversal momenta $q$. The multimode photonic action analogous to Eq.~\eqref{eq:generic_action} can be written in momentum/frequency space as
\begin{align}
    \mathcal{S}_\text{ph} &= \sum_{\mathbf{q},\zeta} (iq_0 - \Omega_q) |a_{\mathbf{q}\zeta}|^2 - \sum_{\mathbf{q}} \Gamma_0^q \Rea \lambda_q^2 a^*_\mathbf{q} a^*_{-\mathbf{q}}\\
    &+ \frac{1}{\beta L^d}\sum_{\mathbf{q},\mathbf{q}',\mathbf{q}''}\begin{aligned}[t] \lambda_q \lambda_{q'} \big[&\Gamma_1^{qq'} a^*_\mathbf{q}a_{\mathbf{-q}'} + \Gamma_2^{qq'} \Rea a^*_\mathbf{q} a^*_{\mathbf{q}'} \\
    + &\Gamma_3^q \delta_{q,-q'}\big]\bphi_{\mathbf{q''}}\bphi_{\mathbf{q}+\mathbf{q}'-\mathbf{q''}}.\end{aligned}\nonumber
\end{align}
where four-momenta are bold, $a_\mathbf{q} = \sum_{\zeta} a_{\mathbf{q}\zeta}$ and the coupling matrices are given in terms of $\Gamma_0^q = -8 B_3 \Gamma_2^{qq}$, $\Gamma_i^{qq'} = \alpha A_i^{q\zeta q'\zeta'}$, ($i=1,2,3$).

From there, the diagrams form Fig.~\ref{fig:diagrams} can be evaluated again, leading for the quadratic vertex to
\begin{align}
    \mathcal{S}_{\text{ph},\bphi^2} = (C + D)\int d^d x \bphi^2
\end{align}
with
\begin{align}
    C &\propto \frac{1}{\beta L^d} \sum_{\mathbf{q}} \frac{(\lambda^2_q \Gamma_2^{qq})^2}{(iq_0 - \Omega_q) (iq_0 +\Omega_q)} \nonumber\\
    &= \frac{1}{2L^d} \sum_{q} \frac{(\lambda_q^2 \Gamma_2^{qq})^2}{\Omega_q}\nonumber\\
    &\approx \int_{\Lambda} \frac{d^d q}{(2\pi)^d} \frac{(\lambda_q^2 \Gamma_2^{qq})^2}{2\Omega_q},\\
    D &\propto \int_{\Lambda} \frac{d^d q}{(2\pi)^d} \lambda_{q}^2 \Gamma_3^q.
\end{align}
In the last step, when transforming the momentum sum, into an integral, the system size dependence is absorbed into the reciprocal space volume element. The resulting self-energy contribution for a given UV cutoff $\Lambda$ is then in general finite when a non-zero set of modes is taken into account in the thermodynamic limit.

The light-induced interaction,
\begin{align}
    \mathcal{S}_{\text{ph},(\bphi^2)^2} = \frac{1}{\beta L^{d}} \sum_{\mathbf{Qqq}'} C(\mathbf{Q}) (\bphi_\mathbf{q} \cdot \bphi_{\mathbf{q}'}) (\bphi_{\mathbf{q+Q}} \cdot \bphi_{\mathbf{q'-Q}}),
\end{align}
similarly loses a factor of $L^{-d}$ in
\begin{align}
    C(\mathbf{Q}) \propto \int_{\Lambda} \frac{d^d q}{(2\pi)^d} \frac{(\lambda_q \lambda_{Q-q} \Gamma_2^{q,Q-q)})^2 (\Omega_q + \Omega_{Q-q})}{(\Omega_q+\Omega_{Q-q})^2 + Q_0^2}.
\end{align}
We therefore conclude that both the shift of the critical point and the light-induced interaction -- provided it is still sufficiently long-ranged -- should become relevant when a finite-measure set of modes is taken into account in the thermodynamic limit, marking a qualitative difference to the single-mode scenario.
\begin{figure}
	\includegraphics{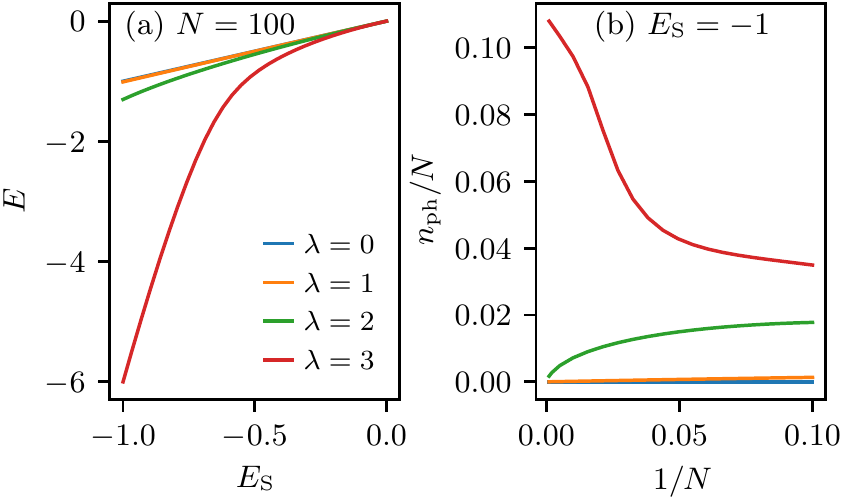}
	\caption{The appearance of false superradiance in the downfolded Peierls model on the diagonally polarized square lattice. In analogy to the main text, $U=200$ and $\Omega/U=0.49$. For $n_\text{ph}^{\max}=200$, the shown results are converged. (a) The cavity-renormalized energy $E$ as a function of bare spin energy $E_\text{S}$ at fixed number of spins. (b) The scaling of the ground state occupation number $n_\text{ph}$ normalized by the system size.}
	\label{fig:variational}
\end{figure}
\section{False superradiance of the downfolded Peierls coupling}
\label{app:superrad}
The Peierls substitution in the original electronic model is gauge invariant under lattice gauge transformations ~\cite{Dmytruk2021}, which should prevent superradiance for a single mode in dipole approximation. Expanding the Peierls phase perturbatively in the light-matter coupling can, however, break this property and lead to false superradiant ground states that would be cut off by higher order terms in the expansion.

The Heisenberg-type model studied in this article is in principle also the product of perturbation theory, since it arises by a perturbative downfolding from the Hubbard model. Even though in this downfolding, the Peierls phase is never expanded and all orders of $\lambda$ are taken into account, the resulting spin Hamiltonian does host a false superradiant phase if $\lambda$ is large enough.

This can be seen in the special case where $\lambda_{i,j}=\text{const}$, i.e. the photon couples exactly the same to all bonds of the magnet, as is the case for the simple square lattice with a diagonally polarized photon. The Hamiltonian in this case becomes
\begin{align}
	H = \Jph H_S + \Omega a^\dag a
\end{align}
which commutes with the spin Hamiltonian $H_S$. We can therefore replace $H_S$ by the pure spin eigenenergies $N E_\text{S}$ and solve the model by diagonalizing the photonic Hilbert space. Without loss of generality, we will work in units in which the ground state energy per spin of $H_S$ is $-1$ so that $E_\text{S}\in [-1,0]$. The cavity coupling does not change the spin wave functions but renormalizes $E_\text{S}$ (Fig.~\ref{fig:variational}(a)). If the renormalized energy becomes nonmonotonous, it can, in principle, change the ground state by reshuffling the original states. However, for the completely negative spectrum of our spin Hamiltonian, this does not happen. The occupation number in the ground state reveals that there is a transition towards macroscopic population at large $\lambda$ (Fig.~\ref{fig:variational}(b)).

Evidently, higher order terms in $t/U$ that are neglected in the perturbative downfolding become important at large occupation numbers, even though the downfolding does take into account all orders of $t^2/U \lambda^{2n}$.

It is important to note that in the main text, we always restrict ourselves to $\lambda$ well below the false superradiant phase, so that $n_\text{ph}$ approaches a small constant.

\section{Universality of the enhancement across lattices}
\label{app:cd_lattice}
In this appendix, we show data similar to Fig.~\ref{fig:scaling_honeycomb} for an alternative dimerized lattice, the square lattice with a columnar arrangement of dimers~\cite{Matsumoto2001}. In the notation of Ref.~\cite{Matsumoto2001}, $J'=\alpha$ is our $J$ and the remaining intradimer bonds are our $\JD$. The cavity mode is polarized perpendicular to the dimers so that it still couples to the critical coupling ratio $J/\JD$ and we tune to the quantum critical point.

\begin{figure}
	\includegraphics{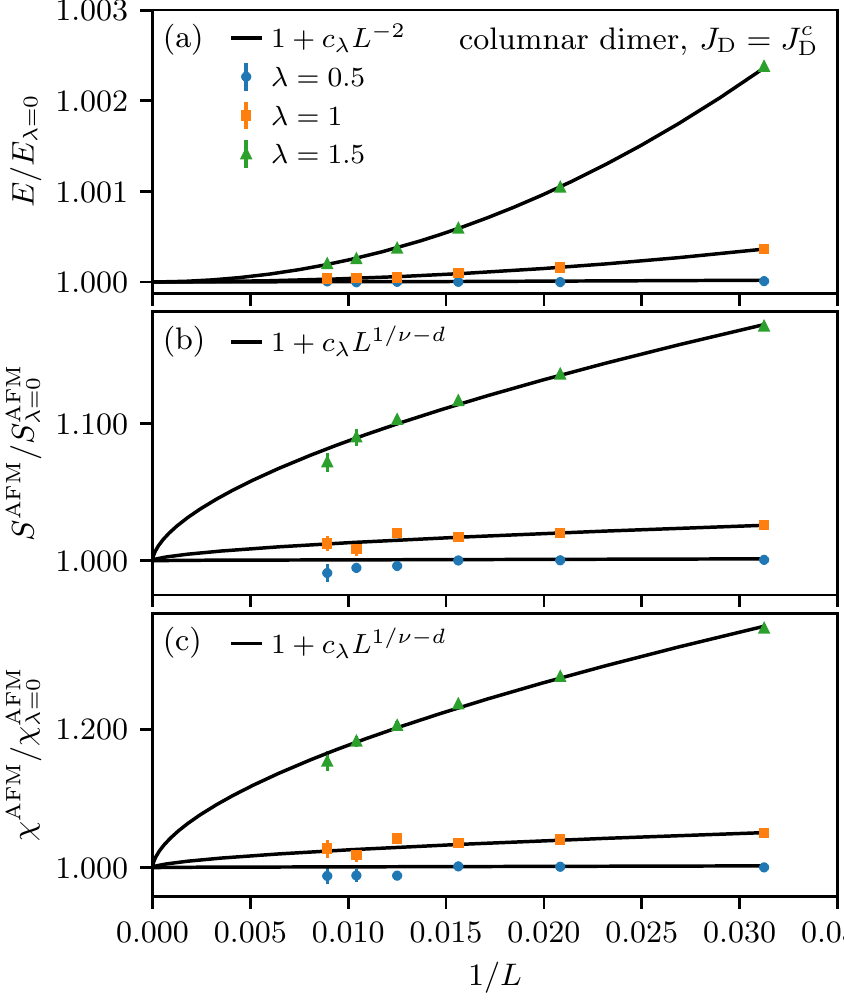}
	\caption{The scaling of different observables in the columnar dimer lattice normalized by their light-matter-decoupled, $\lambda=0$, values. Shown are (a) the energy per spin, $E$, (b) the AFM structure facture corresponding to the ordering pattern of the transition, and (c) the AFM susceptibility, $\chiafm$. The black lines are fits based on a scaling argument in Sec.~\ref{sec:fields}.}
	\label{fig:scaling_columnar_dimer}
\end{figure}

Also in the columnar dimer lattice, the light-matter enhancement of the magnetic structure factor can be described by the exponent $1/\nu-d$ (Fig.~\ref{fig:scaling_columnar_dimer}). This independence on the details of the lattice provides further evidence for the field theoretical argument of section Sec.~\ref{sec:fields}. 
\bibliography{paper.bib}

\begin{thebibliography}{92}%
\makeatletter
\providecommand \@ifxundefined [1]{%
 \@ifx{#1\undefined}
}%
\providecommand \@ifnum [1]{%
 \ifnum #1\expandafter \@firstoftwo
 \else \expandafter \@secondoftwo
 \fi
}%
\providecommand \@ifx [1]{%
 \ifx #1\expandafter \@firstoftwo
 \else \expandafter \@secondoftwo
 \fi
}%
\providecommand \natexlab [1]{#1}%
\providecommand \enquote  [1]{``#1''}%
\providecommand \bibnamefont  [1]{#1}%
\providecommand \bibfnamefont [1]{#1}%
\providecommand \citenamefont [1]{#1}%
\providecommand \href@noop [0]{\@secondoftwo}%
\providecommand \href [0]{\begingroup \@sanitize@url \@href}%
\providecommand \@href[1]{\@@startlink{#1}\@@href}%
\providecommand \@@href[1]{\endgroup#1\@@endlink}%
\providecommand \@sanitize@url [0]{\catcode `\\12\catcode `\$12\catcode
  `\&12\catcode `\#12\catcode `\^12\catcode `\_12\catcode `\%12\relax}%
\providecommand \@@startlink[1]{}%
\providecommand \@@endlink[0]{}%
\providecommand \url  [0]{\begingroup\@sanitize@url \@url }%
\providecommand \@url [1]{\endgroup\@href {#1}{\urlprefix }}%
\providecommand \urlprefix  [0]{URL }%
\providecommand \Eprint [0]{\href }%
\providecommand \doibase [0]{https://doi.org/}%
\providecommand \selectlanguage [0]{\@gobble}%
\providecommand \bibinfo  [0]{\@secondoftwo}%
\providecommand \bibfield  [0]{\@secondoftwo}%
\providecommand \translation [1]{[#1]}%
\providecommand \BibitemOpen [0]{}%
\providecommand \bibitemStop [0]{}%
\providecommand \bibitemNoStop [0]{.\EOS\space}%
\providecommand \EOS [0]{\spacefactor3000\relax}%
\providecommand \BibitemShut  [1]{\csname bibitem#1\endcsname}%
\let\auto@bib@innerbib\@empty
\bibitem [{\citenamefont {Basov}\ \emph {et~al.}(2017)\citenamefont {Basov},
  \citenamefont {Averitt},\ and\ \citenamefont {Hsieh}}]{basov_towards_2017}%
  \BibitemOpen
  \bibfield  {author} {\bibinfo {author} {\bibfnamefont {D.~N.}\ \bibnamefont
  {Basov}}, \bibinfo {author} {\bibfnamefont {R.~D.}\ \bibnamefont {Averitt}},\
  and\ \bibinfo {author} {\bibfnamefont {D.}~\bibnamefont {Hsieh}},\ }\bibfield
   {title} {\bibinfo {title} {Towards properties on demand in quantum
  materials},\ }\href {https://doi.org/10.1038/nmat5017} {\bibfield  {journal}
  {\bibinfo  {journal} {Nat. Mater.}\ }\textbf {\bibinfo {volume} {16}},\
  \bibinfo {pages} {1077} (\bibinfo {year} {2017})}\BibitemShut {NoStop}%
\bibitem [{\citenamefont {Karzig}\ \emph {et~al.}(2015)\citenamefont {Karzig},
  \citenamefont {Bardyn}, \citenamefont {Lindner},\ and\ \citenamefont
  {Refael}}]{karzig_topological_2015}%
  \BibitemOpen
  \bibfield  {author} {\bibinfo {author} {\bibfnamefont {T.}~\bibnamefont
  {Karzig}}, \bibinfo {author} {\bibfnamefont {C.-E.}\ \bibnamefont {Bardyn}},
  \bibinfo {author} {\bibfnamefont {N.~H.}\ \bibnamefont {Lindner}},\ and\
  \bibinfo {author} {\bibfnamefont {G.}~\bibnamefont {Refael}},\ }\bibfield
  {title} {\bibinfo {title} {Topological polaritons},\ }\href
  {https://doi.org/10.1103/PhysRevX.5.031001} {\bibfield  {journal} {\bibinfo
  {journal} {Phys. Rev. X}\ }\textbf {\bibinfo {volume} {5}},\ \bibinfo {pages}
  {031001} (\bibinfo {year} {2015})}\BibitemShut {NoStop}%
\bibitem [{\citenamefont {Claassen}\ \emph {et~al.}(2019)\citenamefont
  {Claassen}, \citenamefont {Kennes}, \citenamefont {Zingl}, \citenamefont
  {Sentef},\ and\ \citenamefont {Rubio}}]{claassen_universal_2019}%
  \BibitemOpen
  \bibfield  {author} {\bibinfo {author} {\bibfnamefont {M.}~\bibnamefont
  {Claassen}}, \bibinfo {author} {\bibfnamefont {D.~M.}\ \bibnamefont
  {Kennes}}, \bibinfo {author} {\bibfnamefont {M.}~\bibnamefont {Zingl}},
  \bibinfo {author} {\bibfnamefont {M.~A.}\ \bibnamefont {Sentef}},\ and\
  \bibinfo {author} {\bibfnamefont {A.}~\bibnamefont {Rubio}},\ }\bibfield
  {title} {\bibinfo {title} {Universal optical control of chiral
  superconductors and {{Majorana}} modes},\ }\href
  {https://doi.org/10.1038/s41567-019-0532-6} {\bibfield  {journal} {\bibinfo
  {journal} {Nat. Phys.}\ }\textbf {\bibinfo {volume} {15}},\ \bibinfo {pages}
  {766} (\bibinfo {year} {2019})}\BibitemShut {NoStop}%
\bibitem [{\citenamefont {Basov}\ \emph {et~al.}(2021)\citenamefont {Basov},
  \citenamefont {Asenjo-Garcia}, \citenamefont {Schuck}, \citenamefont {Zhu},\
  and\ \citenamefont {Rubio}}]{basov_polariton_2021}%
  \BibitemOpen
  \bibfield  {author} {\bibinfo {author} {\bibfnamefont {D.~N.}\ \bibnamefont
  {Basov}}, \bibinfo {author} {\bibfnamefont {A.}~\bibnamefont
  {Asenjo-Garcia}}, \bibinfo {author} {\bibfnamefont {P.~J.}\ \bibnamefont
  {Schuck}}, \bibinfo {author} {\bibfnamefont {X.}~\bibnamefont {Zhu}},\ and\
  \bibinfo {author} {\bibfnamefont {A.}~\bibnamefont {Rubio}},\ }\bibfield
  {title} {\bibinfo {title} {Polariton panorama},\ }\href
  {https://doi.org/10.1515/nanoph-2020-0449} {\bibfield  {journal} {\bibinfo
  {journal} {Nanophotonics}\ }\textbf {\bibinfo {volume} {10}},\ \bibinfo
  {pages} {549} (\bibinfo {year} {2021})}\BibitemShut {NoStop}%
\bibitem [{\citenamefont {Disa}\ \emph {et~al.}(2021)\citenamefont {Disa},
  \citenamefont {Nova},\ and\ \citenamefont
  {Cavalleri}}]{disa_engineering_2021}%
  \BibitemOpen
  \bibfield  {author} {\bibinfo {author} {\bibfnamefont {A.~S.}\ \bibnamefont
  {Disa}}, \bibinfo {author} {\bibfnamefont {T.~F.}\ \bibnamefont {Nova}},\
  and\ \bibinfo {author} {\bibfnamefont {A.}~\bibnamefont {Cavalleri}},\
  }\bibfield  {title} {\bibinfo {title} {Engineering crystal structures with
  light},\ }\href {https://doi.org/10.1038/s41567-021-01366-1} {\bibfield
  {journal} {\bibinfo  {journal} {Nat. Phys.}\ }\textbf {\bibinfo {volume}
  {17}},\ \bibinfo {pages} {1087} (\bibinfo {year} {2021})}\BibitemShut
  {NoStop}%
\bibitem [{\citenamefont {Bloch}\ \emph {et~al.}(2022)\citenamefont {Bloch},
  \citenamefont {Cavalleri}, \citenamefont {Galitski}, \citenamefont {Hafezi},\
  and\ \citenamefont {Rubio}}]{bloch_strongly_2022}%
  \BibitemOpen
  \bibfield  {author} {\bibinfo {author} {\bibfnamefont {J.}~\bibnamefont
  {Bloch}}, \bibinfo {author} {\bibfnamefont {A.}~\bibnamefont {Cavalleri}},
  \bibinfo {author} {\bibfnamefont {V.}~\bibnamefont {Galitski}}, \bibinfo
  {author} {\bibfnamefont {M.}~\bibnamefont {Hafezi}},\ and\ \bibinfo {author}
  {\bibfnamefont {A.}~\bibnamefont {Rubio}},\ }\bibfield  {title} {\bibinfo
  {title} {Strongly correlated electron–photon systems},\ }\href
  {https://doi.org/10.1038/s41586-022-04726-w} {\bibfield  {journal} {\bibinfo
  {journal} {Nature}\ }\textbf {\bibinfo {volume} {606}},\ \bibinfo {pages}
  {41} (\bibinfo {year} {2022})}\BibitemShut {NoStop}%
\bibitem [{\citenamefont {Kockum}\ \emph {et~al.}(2019)\citenamefont {Kockum},
  \citenamefont {Miranowicz}, \citenamefont {Liberato}, \citenamefont
  {Savasta},\ and\ \citenamefont {Nori}}]{Kockum2019}%
  \BibitemOpen
  \bibfield  {author} {\bibinfo {author} {\bibfnamefont {A.~F.}\ \bibnamefont
  {Kockum}}, \bibinfo {author} {\bibfnamefont {A.}~\bibnamefont {Miranowicz}},
  \bibinfo {author} {\bibfnamefont {S.~D.}\ \bibnamefont {Liberato}}, \bibinfo
  {author} {\bibfnamefont {S.}~\bibnamefont {Savasta}},\ and\ \bibinfo {author}
  {\bibfnamefont {F.}~\bibnamefont {Nori}},\ }\bibfield  {title} {\bibinfo
  {title} {Ultrastrong coupling between light and matter},\ }\href
  {https://doi.org/10.1038/s42254-018-0006-2} {\bibfield  {journal} {\bibinfo
  {journal} {Nat. Rev. Phys.}\ }\textbf {\bibinfo {volume} {1}},\ \bibinfo
  {pages} {19} (\bibinfo {year} {2019})}\BibitemShut {NoStop}%
\bibitem [{\citenamefont {H{\"u}bener}\ \emph {et~al.}(2021)\citenamefont
  {H{\"u}bener}, \citenamefont {De~Giovannini}, \citenamefont {Sch{\"a}fer},
  \citenamefont {Andberger}, \citenamefont {Ruggenthaler}, \citenamefont
  {Faist},\ and\ \citenamefont {Rubio}}]{Hubener2021}%
  \BibitemOpen
  \bibfield  {author} {\bibinfo {author} {\bibfnamefont {H.}~\bibnamefont
  {H{\"u}bener}}, \bibinfo {author} {\bibfnamefont {U.}~\bibnamefont
  {De~Giovannini}}, \bibinfo {author} {\bibfnamefont {C.}~\bibnamefont
  {Sch{\"a}fer}}, \bibinfo {author} {\bibfnamefont {J.}~\bibnamefont
  {Andberger}}, \bibinfo {author} {\bibfnamefont {M.}~\bibnamefont
  {Ruggenthaler}}, \bibinfo {author} {\bibfnamefont {J.}~\bibnamefont
  {Faist}},\ and\ \bibinfo {author} {\bibfnamefont {A.}~\bibnamefont {Rubio}},\
  }\bibfield  {title} {\bibinfo {title} {Engineering quantum materials with
  chiral optical cavities},\ }\href
  {https://doi.org/10.1038/s41563-020-00801-7} {\bibfield  {journal} {\bibinfo
  {journal} {Nat. Mater.}\ }\textbf {\bibinfo {volume} {20}},\ \bibinfo {pages}
  {438} (\bibinfo {year} {2021})}\BibitemShut {NoStop}%
\bibitem [{\citenamefont {Schlawin}\ \emph {et~al.}(2022)\citenamefont
  {Schlawin}, \citenamefont {Kennes},\ and\ \citenamefont
  {Sentef}}]{Schlawin2022}%
  \BibitemOpen
  \bibfield  {author} {\bibinfo {author} {\bibfnamefont {F.}~\bibnamefont
  {Schlawin}}, \bibinfo {author} {\bibfnamefont {D.~M.}\ \bibnamefont
  {Kennes}},\ and\ \bibinfo {author} {\bibfnamefont {M.~A.}\ \bibnamefont
  {Sentef}},\ }\bibfield  {title} {\bibinfo {title} {Cavity quantum
  materials},\ }\href {https://doi.org/10.1063/5.0083825} {\bibfield  {journal}
  {\bibinfo  {journal} {Appl. Phys. Rev.}\ }\textbf {\bibinfo {volume} {9}},\
  \bibinfo {pages} {011312} (\bibinfo {year} {2022})}\BibitemShut {NoStop}%
\bibitem [{\citenamefont {Hutchison}\ \emph {et~al.}(2012)\citenamefont
  {Hutchison}, \citenamefont {Schwartz}, \citenamefont {Genet}, \citenamefont
  {Devaux},\ and\ \citenamefont {Ebbesen}}]{Hutchison2012}%
  \BibitemOpen
  \bibfield  {author} {\bibinfo {author} {\bibfnamefont {J.~A.}\ \bibnamefont
  {Hutchison}}, \bibinfo {author} {\bibfnamefont {T.}~\bibnamefont {Schwartz}},
  \bibinfo {author} {\bibfnamefont {C.}~\bibnamefont {Genet}}, \bibinfo
  {author} {\bibfnamefont {E.}~\bibnamefont {Devaux}},\ and\ \bibinfo {author}
  {\bibfnamefont {T.~W.}\ \bibnamefont {Ebbesen}},\ }\bibfield  {title}
  {\bibinfo {title} {Modifying chemical landscapes by coupling to vacuum
  fields},\ }\href {https://doi.org/https://doi.org/10.1002/anie.201107033}
  {\bibfield  {journal} {\bibinfo  {journal} {Angew. Chem. Int. Ed.}\ }\textbf
  {\bibinfo {volume} {51}},\ \bibinfo {pages} {1592} (\bibinfo {year}
  {2012})}\BibitemShut {NoStop}%
\bibitem [{\citenamefont {Galego}\ \emph {et~al.}(2015)\citenamefont {Galego},
  \citenamefont {Garcia-Vidal},\ and\ \citenamefont {Feist}}]{Galego2015}%
  \BibitemOpen
  \bibfield  {author} {\bibinfo {author} {\bibfnamefont {J.}~\bibnamefont
  {Galego}}, \bibinfo {author} {\bibfnamefont {F.~J.}\ \bibnamefont
  {Garcia-Vidal}},\ and\ \bibinfo {author} {\bibfnamefont {J.}~\bibnamefont
  {Feist}},\ }\bibfield  {title} {\bibinfo {title} {Cavity-induced
  modifications of molecular structure in the strong-coupling regime},\ }\href
  {https://doi.org/10.1103/PhysRevX.5.041022} {\bibfield  {journal} {\bibinfo
  {journal} {Phys. Rev. X}\ }\textbf {\bibinfo {volume} {5}},\ \bibinfo {pages}
  {041022} (\bibinfo {year} {2015})}\BibitemShut {NoStop}%
\bibitem [{\citenamefont {Ebbesen}(2016)}]{Ebbesen2016}%
  \BibitemOpen
  \bibfield  {author} {\bibinfo {author} {\bibfnamefont {T.~W.}\ \bibnamefont
  {Ebbesen}},\ }\bibfield  {title} {\bibinfo {title} {Hybrid light--matter
  states in a molecular and material science perspective},\ }\href
  {https://doi.org/10.1021/acs.accounts.6b00295} {\bibfield  {journal}
  {\bibinfo  {journal} {Acc. Chem. Res.}\ }\textbf {\bibinfo {volume} {49}},\
  \bibinfo {pages} {2403} (\bibinfo {year} {2016})}\BibitemShut {NoStop}%
\bibitem [{\citenamefont {Feist}\ \emph {et~al.}(2018)\citenamefont {Feist},
  \citenamefont {Galego},\ and\ \citenamefont {Garcia-Vidal}}]{Feist2018}%
  \BibitemOpen
  \bibfield  {author} {\bibinfo {author} {\bibfnamefont {J.}~\bibnamefont
  {Feist}}, \bibinfo {author} {\bibfnamefont {J.}~\bibnamefont {Galego}},\ and\
  \bibinfo {author} {\bibfnamefont {F.~J.}\ \bibnamefont {Garcia-Vidal}},\
  }\bibfield  {title} {\bibinfo {title} {Polaritonic chemistry with organic
  molecules},\ }\href {https://doi.org/10.1021/acsphotonics.7b00680} {\bibfield
   {journal} {\bibinfo  {journal} {ACS Photonics}\ }\textbf {\bibinfo {volume}
  {5}},\ \bibinfo {pages} {205} (\bibinfo {year} {2018})}\BibitemShut {NoStop}%
\bibitem [{\citenamefont {F.~Ribeiro}\ \emph {et~al.}(2018)\citenamefont
  {F.~Ribeiro}, \citenamefont {A.~Martínez-Martínez}, \citenamefont {Du},
  \citenamefont {Campos-Gonzalez-Angulo},\ and\ \citenamefont
  {Yuen-Zhou}}]{fribeiro_polariton_2018}%
  \BibitemOpen
  \bibfield  {author} {\bibinfo {author} {\bibfnamefont {R.}~\bibnamefont
  {F.~Ribeiro}}, \bibinfo {author} {\bibfnamefont {L.}~\bibnamefont
  {A.~Martínez-Martínez}}, \bibinfo {author} {\bibfnamefont {M.}~\bibnamefont
  {Du}}, \bibinfo {author} {\bibfnamefont {J.}~\bibnamefont
  {Campos-Gonzalez-Angulo}},\ and\ \bibinfo {author} {\bibfnamefont
  {J.}~\bibnamefont {Yuen-Zhou}},\ }\bibfield  {title} {\bibinfo {title}
  {Polariton chemistry: controlling molecular dynamics with optical cavities},\
  }\href {https://doi.org/10.1039/C8SC01043A} {\bibfield  {journal} {\bibinfo
  {journal} {Chem. Sci.}\ }\textbf {\bibinfo {volume} {9}},\ \bibinfo {pages}
  {6325} (\bibinfo {year} {2018})}\BibitemShut {NoStop}%
\bibitem [{\citenamefont {Schäfer}\ \emph {et~al.}(2019)\citenamefont
  {Schäfer}, \citenamefont {Ruggenthaler}, \citenamefont {Appel},\ and\
  \citenamefont {Rubio}}]{schafer_modification_2019}%
  \BibitemOpen
  \bibfield  {author} {\bibinfo {author} {\bibfnamefont {C.}~\bibnamefont
  {Schäfer}}, \bibinfo {author} {\bibfnamefont {M.}~\bibnamefont
  {Ruggenthaler}}, \bibinfo {author} {\bibfnamefont {H.}~\bibnamefont
  {Appel}},\ and\ \bibinfo {author} {\bibfnamefont {A.}~\bibnamefont {Rubio}},\
  }\bibfield  {title} {\bibinfo {title} {Modification of excitation and charge
  transfer in cavity quantum-electrodynamical chemistry},\ }\href
  {https://doi.org/10.1073/pnas.1814178116} {\bibfield  {journal} {\bibinfo
  {journal} {Proc. Natl. Acad. Sci. U.S.A.}\ }\textbf {\bibinfo {volume}
  {116}},\ \bibinfo {pages} {4883} (\bibinfo {year} {2019})}\BibitemShut
  {NoStop}%
\bibitem [{\citenamefont {Sidler}\ \emph {et~al.}(2021)\citenamefont {Sidler},
  \citenamefont {Schäfer}, \citenamefont {Ruggenthaler},\ and\ \citenamefont
  {Rubio}}]{sidler_polaritonic_2021}%
  \BibitemOpen
  \bibfield  {author} {\bibinfo {author} {\bibfnamefont {D.}~\bibnamefont
  {Sidler}}, \bibinfo {author} {\bibfnamefont {C.}~\bibnamefont {Schäfer}},
  \bibinfo {author} {\bibfnamefont {M.}~\bibnamefont {Ruggenthaler}},\ and\
  \bibinfo {author} {\bibfnamefont {A.}~\bibnamefont {Rubio}},\ }\bibfield
  {title} {\bibinfo {title} {Polaritonic chemistry: Collective strong coupling
  implies strong local modification of chemical properties},\ }\href
  {https://doi.org/10.1021/acs.jpclett.0c03436} {\bibfield  {journal} {\bibinfo
   {journal} {J. Phys. Chem. Lett.}\ }\textbf {\bibinfo {volume} {12}},\
  \bibinfo {pages} {508} (\bibinfo {year} {2021})}\BibitemShut {NoStop}%
\bibitem [{\citenamefont {Sidler}\ \emph {et~al.}(2022)\citenamefont {Sidler},
  \citenamefont {Ruggenthaler}, \citenamefont {Schäfer}, \citenamefont
  {Ronca},\ and\ \citenamefont {Rubio}}]{Sidler2022}%
  \BibitemOpen
  \bibfield  {author} {\bibinfo {author} {\bibfnamefont {D.}~\bibnamefont
  {Sidler}}, \bibinfo {author} {\bibfnamefont {M.}~\bibnamefont
  {Ruggenthaler}}, \bibinfo {author} {\bibfnamefont {C.}~\bibnamefont
  {Schäfer}}, \bibinfo {author} {\bibfnamefont {E.}~\bibnamefont {Ronca}},\
  and\ \bibinfo {author} {\bibfnamefont {A.}~\bibnamefont {Rubio}},\ }\bibfield
   {title} {\bibinfo {title} {A perspective on ab initio modeling of
  polaritonic chemistry: The role of non-equilibrium effects and quantum
  collectivity},\ }\href {https://doi.org/10.1063/5.0094956} {\bibfield
  {journal} {\bibinfo  {journal} {J. Chem. Phys.}\ }\textbf {\bibinfo {volume}
  {156}},\ \bibinfo {pages} {230901} (\bibinfo {year} {2022})}\BibitemShut
  {NoStop}%
\bibitem [{\citenamefont {Li}\ \emph {et~al.}(2022)\citenamefont {Li},
  \citenamefont {Cui}, \citenamefont {Subotnik},\ and\ \citenamefont
  {Nitzan}}]{li_molecular_2022}%
  \BibitemOpen
  \bibfield  {author} {\bibinfo {author} {\bibfnamefont {T.~E.}\ \bibnamefont
  {Li}}, \bibinfo {author} {\bibfnamefont {B.}~\bibnamefont {Cui}}, \bibinfo
  {author} {\bibfnamefont {J.~E.}\ \bibnamefont {Subotnik}},\ and\ \bibinfo
  {author} {\bibfnamefont {A.}~\bibnamefont {Nitzan}},\ }\bibfield  {title}
  {\bibinfo {title} {Molecular polaritonics: Chemical dynamics under strong
  light–matter coupling},\ }\href
  {https://doi.org/10.1146/annurev-physchem-090519-042621} {\bibfield
  {journal} {\bibinfo  {journal} {Annu. Rev. Phys. Chem.}\ }\textbf {\bibinfo
  {volume} {73}},\ \bibinfo {pages} {43} (\bibinfo {year} {2022})}\BibitemShut
  {NoStop}%
\bibitem [{\citenamefont {Genet}\ \emph {et~al.}(2021)\citenamefont {Genet},
  \citenamefont {Faist},\ and\ \citenamefont {Ebbesen}}]{genet_inducing_2021}%
  \BibitemOpen
  \bibfield  {author} {\bibinfo {author} {\bibfnamefont {C.}~\bibnamefont
  {Genet}}, \bibinfo {author} {\bibfnamefont {J.}~\bibnamefont {Faist}},\ and\
  \bibinfo {author} {\bibfnamefont {T.~W.}\ \bibnamefont {Ebbesen}},\
  }\bibfield  {title} {\bibinfo {title} {Inducing new material properties with
  hybrid light–matter states},\ }\href {https://doi.org/10.1063/PT.3.4749}
  {\bibfield  {journal} {\bibinfo  {journal} {Phys. Today}\ }\textbf {\bibinfo
  {volume} {74}},\ \bibinfo {pages} {42} (\bibinfo {year} {2021})}\BibitemShut
  {NoStop}%
\bibitem [{\citenamefont {D'Alessio}\ and\ \citenamefont
  {Rigol}(2014)}]{DAlessio2014}%
  \BibitemOpen
  \bibfield  {author} {\bibinfo {author} {\bibfnamefont {L.}~\bibnamefont
  {D'Alessio}}\ and\ \bibinfo {author} {\bibfnamefont {M.}~\bibnamefont
  {Rigol}},\ }\bibfield  {title} {\bibinfo {title} {Long-time behavior of
  isolated periodically driven interacting lattice systems},\ }\href
  {https://doi.org/10.1103/PhysRevX.4.041048} {\bibfield  {journal} {\bibinfo
  {journal} {Phys. Rev. X}\ }\textbf {\bibinfo {volume} {4}},\ \bibinfo {pages}
  {041048} (\bibinfo {year} {2014})}\BibitemShut {NoStop}%
\bibitem [{\citenamefont {Lazarides}\ \emph {et~al.}(2014)\citenamefont
  {Lazarides}, \citenamefont {Das},\ and\ \citenamefont
  {Moessner}}]{Lazarides2014}%
  \BibitemOpen
  \bibfield  {author} {\bibinfo {author} {\bibfnamefont {A.}~\bibnamefont
  {Lazarides}}, \bibinfo {author} {\bibfnamefont {A.}~\bibnamefont {Das}},\
  and\ \bibinfo {author} {\bibfnamefont {R.}~\bibnamefont {Moessner}},\
  }\bibfield  {title} {\bibinfo {title} {Equilibrium states of generic quantum
  systems subject to periodic driving},\ }\href
  {https://doi.org/10.1103/PhysRevE.90.012110} {\bibfield  {journal} {\bibinfo
  {journal} {Phys. Rev. E}\ }\textbf {\bibinfo {volume} {90}},\ \bibinfo
  {pages} {012110} (\bibinfo {year} {2014})}\BibitemShut {NoStop}%
\bibitem [{\citenamefont {Kennes}\ \emph {et~al.}(2018)\citenamefont {Kennes},
  \citenamefont {de~la Torre}, \citenamefont {Ron}, \citenamefont {Hsieh},\
  and\ \citenamefont {Millis}}]{Kennes2018}%
  \BibitemOpen
  \bibfield  {author} {\bibinfo {author} {\bibfnamefont {D.~M.}\ \bibnamefont
  {Kennes}}, \bibinfo {author} {\bibfnamefont {A.}~\bibnamefont {de~la Torre}},
  \bibinfo {author} {\bibfnamefont {A.}~\bibnamefont {Ron}}, \bibinfo {author}
  {\bibfnamefont {D.}~\bibnamefont {Hsieh}},\ and\ \bibinfo {author}
  {\bibfnamefont {A.~J.}\ \bibnamefont {Millis}},\ }\bibfield  {title}
  {\bibinfo {title} {Floquet engineering in quantum chains},\ }\href
  {https://doi.org/10.1103/PhysRevLett.120.127601} {\bibfield  {journal}
  {\bibinfo  {journal} {Phys. Rev. Lett.}\ }\textbf {\bibinfo {volume} {120}},\
  \bibinfo {pages} {127601} (\bibinfo {year} {2018})}\BibitemShut {NoStop}%
\bibitem [{\citenamefont {Sentef}\ \emph {et~al.}(2020)\citenamefont {Sentef},
  \citenamefont {Li}, \citenamefont {K\"unzel},\ and\ \citenamefont
  {Eckstein}}]{Sentef2020}%
  \BibitemOpen
  \bibfield  {author} {\bibinfo {author} {\bibfnamefont {M.~A.}\ \bibnamefont
  {Sentef}}, \bibinfo {author} {\bibfnamefont {J.}~\bibnamefont {Li}}, \bibinfo
  {author} {\bibfnamefont {F.}~\bibnamefont {K\"unzel}},\ and\ \bibinfo
  {author} {\bibfnamefont {M.}~\bibnamefont {Eckstein}},\ }\bibfield  {title}
  {\bibinfo {title} {Quantum to classical crossover of {F}loquet engineering in
  correlated quantum systems},\ }\href
  {https://doi.org/10.1103/PhysRevResearch.2.033033} {\bibfield  {journal}
  {\bibinfo  {journal} {Phys. Rev. Res.}\ }\textbf {\bibinfo {volume} {2}},\
  \bibinfo {pages} {033033} (\bibinfo {year} {2020})}\BibitemShut {NoStop}%
\bibitem [{\citenamefont {Kibis}\ \emph {et~al.}(2011)\citenamefont {Kibis},
  \citenamefont {Kyriienko},\ and\ \citenamefont {Shelykh}}]{Kibis2011}%
  \BibitemOpen
  \bibfield  {author} {\bibinfo {author} {\bibfnamefont {O.~V.}\ \bibnamefont
  {Kibis}}, \bibinfo {author} {\bibfnamefont {O.}~\bibnamefont {Kyriienko}},\
  and\ \bibinfo {author} {\bibfnamefont {I.~A.}\ \bibnamefont {Shelykh}},\
  }\bibfield  {title} {\bibinfo {title} {Band gap in graphene induced by vacuum
  fluctuations},\ }\href {https://doi.org/10.1103/PhysRevB.84.195413}
  {\bibfield  {journal} {\bibinfo  {journal} {Phys. Rev. B}\ }\textbf {\bibinfo
  {volume} {84}},\ \bibinfo {pages} {195413} (\bibinfo {year}
  {2011})}\BibitemShut {NoStop}%
\bibitem [{\citenamefont {Wang}\ \emph {et~al.}(2019)\citenamefont {Wang},
  \citenamefont {Ronca},\ and\ \citenamefont {Sentef}}]{Sentef2019}%
  \BibitemOpen
  \bibfield  {author} {\bibinfo {author} {\bibfnamefont {X.}~\bibnamefont
  {Wang}}, \bibinfo {author} {\bibfnamefont {E.}~\bibnamefont {Ronca}},\ and\
  \bibinfo {author} {\bibfnamefont {M.~A.}\ \bibnamefont {Sentef}},\ }\bibfield
   {title} {\bibinfo {title} {Cavity quantum electrodynamical {C}hern
  insulator: Towards light-induced quantized anomalous {H}all effect in
  graphene},\ }\href {https://doi.org/10.1103/PhysRevB.99.235156} {\bibfield
  {journal} {\bibinfo  {journal} {Phys. Rev. B}\ }\textbf {\bibinfo {volume}
  {99}},\ \bibinfo {pages} {235156} (\bibinfo {year} {2019})}\BibitemShut
  {NoStop}%
\bibitem [{\citenamefont {Dmytruk}\ and\ \citenamefont
  {Schirò}(2022)}]{Dmytruk2022}%
  \BibitemOpen
  \bibfield  {author} {\bibinfo {author} {\bibfnamefont {O.}~\bibnamefont
  {Dmytruk}}\ and\ \bibinfo {author} {\bibfnamefont {M.}~\bibnamefont
  {Schirò}},\ }\href@noop {} {\bibinfo {title} {Controlling topological phases
  of matter with quantum light}} (\bibinfo {year} {2022}),\ \Eprint
  {https://arxiv.org/abs/2204.05922} {arXiv:2204.05922 [cond-mat.mes-hall]}
  \BibitemShut {NoStop}%
\bibitem [{\citenamefont {Ashida}\ \emph {et~al.}(2020)\citenamefont {Ashida},
  \citenamefont {\ifmmode \dot{I}\else \.{I}\fi{}mamo\ifmmode~\breve{g}\else
  \u{g}\fi{}lu}, \citenamefont {Faist}, \citenamefont {Jaksch}, \citenamefont
  {Cavalleri},\ and\ \citenamefont {Demler}}]{Ashida2020}%
  \BibitemOpen
  \bibfield  {author} {\bibinfo {author} {\bibfnamefont {Y.}~\bibnamefont
  {Ashida}}, \bibinfo {author} {\bibfnamefont {A.~m.~c.}\ \bibnamefont
  {\ifmmode \dot{I}\else \.{I}\fi{}mamo\ifmmode~\breve{g}\else \u{g}\fi{}lu}},
  \bibinfo {author} {\bibfnamefont {J.}~\bibnamefont {Faist}}, \bibinfo
  {author} {\bibfnamefont {D.}~\bibnamefont {Jaksch}}, \bibinfo {author}
  {\bibfnamefont {A.}~\bibnamefont {Cavalleri}},\ and\ \bibinfo {author}
  {\bibfnamefont {E.}~\bibnamefont {Demler}},\ }\bibfield  {title} {\bibinfo
  {title} {Quantum electrodynamic control of matter: Cavity-enhanced
  ferroelectric phase transition},\ }\href
  {https://doi.org/10.1103/PhysRevX.10.041027} {\bibfield  {journal} {\bibinfo
  {journal} {Phys. Rev. X}\ }\textbf {\bibinfo {volume} {10}},\ \bibinfo
  {pages} {041027} (\bibinfo {year} {2020})}\BibitemShut {NoStop}%
\bibitem [{\citenamefont {Latini}\ \emph {et~al.}(2021)\citenamefont {Latini},
  \citenamefont {Shin}, \citenamefont {Sato}, \citenamefont {Schäfer},
  \citenamefont {Giovannini}, \citenamefont {Hübener},\ and\ \citenamefont
  {Rubio}}]{Latini2021}%
  \BibitemOpen
  \bibfield  {author} {\bibinfo {author} {\bibfnamefont {S.}~\bibnamefont
  {Latini}}, \bibinfo {author} {\bibfnamefont {D.}~\bibnamefont {Shin}},
  \bibinfo {author} {\bibfnamefont {S.~A.}\ \bibnamefont {Sato}}, \bibinfo
  {author} {\bibfnamefont {C.}~\bibnamefont {Schäfer}}, \bibinfo {author}
  {\bibfnamefont {U.~D.}\ \bibnamefont {Giovannini}}, \bibinfo {author}
  {\bibfnamefont {H.}~\bibnamefont {Hübener}},\ and\ \bibinfo {author}
  {\bibfnamefont {A.}~\bibnamefont {Rubio}},\ }\bibfield  {title} {\bibinfo
  {title} {The ferroelectric photo ground state of {SrTiO$_3$}: Cavity
  materials engineering},\ }\href {https://doi.org/10.1073/pnas.2105618118}
  {\bibfield  {journal} {\bibinfo  {journal} {Proc. Natl. Acad. Sci. U.S.A.}\
  }\textbf {\bibinfo {volume} {118}},\ \bibinfo {pages} {e2105618118} (\bibinfo
  {year} {2021})}\BibitemShut {NoStop}%
\bibitem [{\citenamefont {Mazza}\ and\ \citenamefont
  {Georges}(2019)}]{Mazza2019}%
  \BibitemOpen
  \bibfield  {author} {\bibinfo {author} {\bibfnamefont {G.}~\bibnamefont
  {Mazza}}\ and\ \bibinfo {author} {\bibfnamefont {A.}~\bibnamefont
  {Georges}},\ }\bibfield  {title} {\bibinfo {title} {Superradiant quantum
  materials},\ }\href {https://doi.org/10.1103/PhysRevLett.122.017401}
  {\bibfield  {journal} {\bibinfo  {journal} {Phys. Rev. Lett.}\ }\textbf
  {\bibinfo {volume} {122}},\ \bibinfo {pages} {017401} (\bibinfo {year}
  {2019})}\BibitemShut {NoStop}%
\bibitem [{\citenamefont {{Vi{\~n}as Bostr{\"o}m}}\ \emph
  {et~al.}(2022)\citenamefont {{Vi{\~n}as Bostr{\"o}m}}, \citenamefont
  {Sriram}, \citenamefont {Claassen},\ and\ \citenamefont
  {Rubio}}]{Bostrom2022}%
  \BibitemOpen
  \bibfield  {author} {\bibinfo {author} {\bibfnamefont {E.}~\bibnamefont
  {{Vi{\~n}as Bostr{\"o}m}}}, \bibinfo {author} {\bibfnamefont
  {A.}~\bibnamefont {Sriram}}, \bibinfo {author} {\bibfnamefont
  {M.}~\bibnamefont {Claassen}},\ and\ \bibinfo {author} {\bibfnamefont
  {A.}~\bibnamefont {Rubio}},\ }\bibfield  {title} {\bibinfo {title}
  {Controlling the magnetic state of the proximate quantum spin liquid
  $\alpha$-{RuCl}$_3$ with an optical cavity},\ }\href@noop {} {\bibfield
  {journal} {\bibinfo  {journal} {arXiv:2211.07247}\ } (\bibinfo {year}
  {2022})}\BibitemShut {NoStop}%
\bibitem [{\citenamefont {Chiocchetta}\ \emph {et~al.}(2021)\citenamefont
  {Chiocchetta}, \citenamefont {Kiese}, \citenamefont {Zelle}, \citenamefont
  {Piazza},\ and\ \citenamefont {Diehl}}]{Chiocchetta2021}%
  \BibitemOpen
  \bibfield  {author} {\bibinfo {author} {\bibfnamefont {A.}~\bibnamefont
  {Chiocchetta}}, \bibinfo {author} {\bibfnamefont {D.}~\bibnamefont {Kiese}},
  \bibinfo {author} {\bibfnamefont {C.~P.}\ \bibnamefont {Zelle}}, \bibinfo
  {author} {\bibfnamefont {F.}~\bibnamefont {Piazza}},\ and\ \bibinfo {author}
  {\bibfnamefont {S.}~\bibnamefont {Diehl}},\ }\bibfield  {title} {\bibinfo
  {title} {Cavity-induced quantum spin liquids},\ }\href
  {https://doi.org/10.1038/s41467-021-26076-3} {\bibfield  {journal} {\bibinfo
  {journal} {Nat. Commun.}\ }\textbf {\bibinfo {volume} {12}},\ \bibinfo
  {pages} {5901} (\bibinfo {year} {2021})}\BibitemShut {NoStop}%
\bibitem [{\citenamefont {Laussy}\ \emph {et~al.}(2010)\citenamefont {Laussy},
  \citenamefont {Kavokin},\ and\ \citenamefont {Shelykh}}]{Laussy2010}%
  \BibitemOpen
  \bibfield  {author} {\bibinfo {author} {\bibfnamefont {F.~P.}\ \bibnamefont
  {Laussy}}, \bibinfo {author} {\bibfnamefont {A.~V.}\ \bibnamefont
  {Kavokin}},\ and\ \bibinfo {author} {\bibfnamefont {I.~A.}\ \bibnamefont
  {Shelykh}},\ }\bibfield  {title} {\bibinfo {title} {Exciton-polariton
  mediated superconductivity},\ }\href
  {https://doi.org/10.1103/PhysRevLett.104.106402} {\bibfield  {journal}
  {\bibinfo  {journal} {Phys. Rev. Lett.}\ }\textbf {\bibinfo {volume} {104}},\
  \bibinfo {pages} {106402} (\bibinfo {year} {2010})}\BibitemShut {NoStop}%
\bibitem [{\citenamefont {Laplace}\ \emph {et~al.}(2016)\citenamefont
  {Laplace}, \citenamefont {Fernandez-Pena}, \citenamefont {Gariglio},
  \citenamefont {Triscone},\ and\ \citenamefont {Cavalleri}}]{Laplace2016}%
  \BibitemOpen
  \bibfield  {author} {\bibinfo {author} {\bibfnamefont {Y.}~\bibnamefont
  {Laplace}}, \bibinfo {author} {\bibfnamefont {S.}~\bibnamefont
  {Fernandez-Pena}}, \bibinfo {author} {\bibfnamefont {S.}~\bibnamefont
  {Gariglio}}, \bibinfo {author} {\bibfnamefont {J.~M.}\ \bibnamefont
  {Triscone}},\ and\ \bibinfo {author} {\bibfnamefont {A.}~\bibnamefont
  {Cavalleri}},\ }\bibfield  {title} {\bibinfo {title} {Proposed cavity
  josephson plasmonics with complex-oxide heterostructures},\ }\href
  {https://doi.org/10.1103/PhysRevB.93.075152} {\bibfield  {journal} {\bibinfo
  {journal} {Phys. Rev. B}\ }\textbf {\bibinfo {volume} {93}},\ \bibinfo
  {pages} {075152} (\bibinfo {year} {2016})}\BibitemShut {NoStop}%
\bibitem [{\citenamefont {Sentef}\ \emph {et~al.}(2018)\citenamefont {Sentef},
  \citenamefont {Ruggenthaler},\ and\ \citenamefont {Rubio}}]{Sentef2018}%
  \BibitemOpen
  \bibfield  {author} {\bibinfo {author} {\bibfnamefont {M.~A.}\ \bibnamefont
  {Sentef}}, \bibinfo {author} {\bibfnamefont {M.}~\bibnamefont
  {Ruggenthaler}},\ and\ \bibinfo {author} {\bibfnamefont {A.}~\bibnamefont
  {Rubio}},\ }\bibfield  {title} {\bibinfo {title} {Cavity
  quantum-electrodynamical polaritonically enhanced electron-phonon coupling
  and its influence on superconductivity},\ }\href
  {https://doi.org/10.1126/sciadv.aau6969} {\bibfield  {journal} {\bibinfo
  {journal} {Sci. Adv.}\ }\textbf {\bibinfo {volume} {4}},\ \bibinfo {pages}
  {eaau6969} (\bibinfo {year} {2018})}\BibitemShut {NoStop}%
\bibitem [{\citenamefont {Hagenm{\"u}ller}\ \emph {et~al.}(2019)\citenamefont
  {Hagenm{\"u}ller}, \citenamefont {Schachenmayer}, \citenamefont {Genet},
  \citenamefont {Ebbesen},\ and\ \citenamefont {Pupillo}}]{Hagenmuller2019}%
  \BibitemOpen
  \bibfield  {author} {\bibinfo {author} {\bibfnamefont {D.}~\bibnamefont
  {Hagenm{\"u}ller}}, \bibinfo {author} {\bibfnamefont {J.}~\bibnamefont
  {Schachenmayer}}, \bibinfo {author} {\bibfnamefont {C.}~\bibnamefont
  {Genet}}, \bibinfo {author} {\bibfnamefont {T.~W.}\ \bibnamefont {Ebbesen}},\
  and\ \bibinfo {author} {\bibfnamefont {G.}~\bibnamefont {Pupillo}},\
  }\bibfield  {title} {\bibinfo {title} {Enhancement of the electron--phonon
  scattering induced by intrinsic surface plasmon--phonon polaritons},\ }\href
  {https://doi.org/10.1021/acsphotonics.9b00268} {\bibfield  {journal}
  {\bibinfo  {journal} {ACS Photonics}\ }\textbf {\bibinfo {volume} {6}},\
  \bibinfo {pages} {1073} (\bibinfo {year} {2019})}\BibitemShut {NoStop}%
\bibitem [{\citenamefont {Gao}\ \emph {et~al.}(2020)\citenamefont {Gao},
  \citenamefont {Schlawin}, \citenamefont {Buzzi}, \citenamefont {Cavalleri},\
  and\ \citenamefont {Jaksch}}]{Gao2020}%
  \BibitemOpen
  \bibfield  {author} {\bibinfo {author} {\bibfnamefont {H.}~\bibnamefont
  {Gao}}, \bibinfo {author} {\bibfnamefont {F.}~\bibnamefont {Schlawin}},
  \bibinfo {author} {\bibfnamefont {M.}~\bibnamefont {Buzzi}}, \bibinfo
  {author} {\bibfnamefont {A.}~\bibnamefont {Cavalleri}},\ and\ \bibinfo
  {author} {\bibfnamefont {D.}~\bibnamefont {Jaksch}},\ }\bibfield  {title}
  {\bibinfo {title} {Photoinduced electron pairing in a driven cavity},\ }\href
  {https://doi.org/10.1103/PhysRevLett.125.053602} {\bibfield  {journal}
  {\bibinfo  {journal} {Phys. Rev. Lett.}\ }\textbf {\bibinfo {volume} {125}},\
  \bibinfo {pages} {053602} (\bibinfo {year} {2020})}\BibitemShut {NoStop}%
\bibitem [{\citenamefont {Chakraborty}\ and\ \citenamefont
  {Piazza}(2021)}]{Chakraborty2021}%
  \BibitemOpen
  \bibfield  {author} {\bibinfo {author} {\bibfnamefont {A.}~\bibnamefont
  {Chakraborty}}\ and\ \bibinfo {author} {\bibfnamefont {F.}~\bibnamefont
  {Piazza}},\ }\bibfield  {title} {\bibinfo {title} {Long-range photon
  fluctuations enhance photon-mediated electron pairing and
  superconductivity},\ }\href {https://doi.org/10.1103/PhysRevLett.127.177002}
  {\bibfield  {journal} {\bibinfo  {journal} {Phys. Rev. Lett.}\ }\textbf
  {\bibinfo {volume} {127}},\ \bibinfo {pages} {177002} (\bibinfo {year}
  {2021})}\BibitemShut {NoStop}%
\bibitem [{\citenamefont {Sachdev}(2011)}]{sachdev_quantum_2011}%
  \BibitemOpen
  \bibfield  {author} {\bibinfo {author} {\bibfnamefont {S.}~\bibnamefont
  {Sachdev}},\ }\href {https://doi.org/10.1017/CBO9780511973765} {\emph
  {\bibinfo {title} {Quantum Phase Transitions}}},\ \bibinfo {edition} {2nd}\
  ed.\ (\bibinfo  {publisher} {Cambridge University Press},\ \bibinfo {address}
  {Cambridge},\ \bibinfo {year} {2011})\BibitemShut {NoStop}%
\bibitem [{\citenamefont {Ruggenthaler}\ \emph {et~al.}(2018)\citenamefont
  {Ruggenthaler}, \citenamefont {Tancogne-Dejean}, \citenamefont {Flick},
  \citenamefont {Appel},\ and\ \citenamefont {Rubio}}]{Ruggenthaler2018}%
  \BibitemOpen
  \bibfield  {author} {\bibinfo {author} {\bibfnamefont {M.}~\bibnamefont
  {Ruggenthaler}}, \bibinfo {author} {\bibfnamefont {N.}~\bibnamefont
  {Tancogne-Dejean}}, \bibinfo {author} {\bibfnamefont {J.}~\bibnamefont
  {Flick}}, \bibinfo {author} {\bibfnamefont {H.}~\bibnamefont {Appel}},\ and\
  \bibinfo {author} {\bibfnamefont {A.}~\bibnamefont {Rubio}},\ }\bibfield
  {title} {\bibinfo {title} {From a quantum-electrodynamical light--matter
  description to novel spectroscopies},\ }\href
  {https://doi.org/10.1038/s41570-018-0118} {\bibfield  {journal} {\bibinfo
  {journal} {Nat. Rev. Chem.}\ }\textbf {\bibinfo {volume} {2}},\ \bibinfo
  {pages} {0118} (\bibinfo {year} {2018})}\BibitemShut {NoStop}%
\bibitem [{\citenamefont {Rokaj}\ \emph {et~al.}(2018)\citenamefont {Rokaj},
  \citenamefont {Welakuh}, \citenamefont {Ruggenthaler},\ and\ \citenamefont
  {Rubio}}]{rokaj_lightmatter_2018}%
  \BibitemOpen
  \bibfield  {author} {\bibinfo {author} {\bibfnamefont {V.}~\bibnamefont
  {Rokaj}}, \bibinfo {author} {\bibfnamefont {D.~M.}\ \bibnamefont {Welakuh}},
  \bibinfo {author} {\bibfnamefont {M.}~\bibnamefont {Ruggenthaler}},\ and\
  \bibinfo {author} {\bibfnamefont {A.}~\bibnamefont {Rubio}},\ }\bibfield
  {title} {\bibinfo {title} {Light–matter interaction in the long-wavelength
  limit: no ground-state without dipole self-energy},\ }\href
  {https://doi.org/10.1088/1361-6455/aa9c99} {\bibfield  {journal} {\bibinfo
  {journal} {J. Phys. B}\ }\textbf {\bibinfo {volume} {51}},\ \bibinfo {pages}
  {034005} (\bibinfo {year} {2018})}\BibitemShut {NoStop}%
\bibitem [{\citenamefont {Schäfer}\ \emph {et~al.}(2020)\citenamefont
  {Schäfer}, \citenamefont {Ruggenthaler}, \citenamefont {Rokaj},\ and\
  \citenamefont {Rubio}}]{schafer_relevance_2020}%
  \BibitemOpen
  \bibfield  {author} {\bibinfo {author} {\bibfnamefont {C.}~\bibnamefont
  {Schäfer}}, \bibinfo {author} {\bibfnamefont {M.}~\bibnamefont
  {Ruggenthaler}}, \bibinfo {author} {\bibfnamefont {V.}~\bibnamefont
  {Rokaj}},\ and\ \bibinfo {author} {\bibfnamefont {A.}~\bibnamefont {Rubio}},\
  }\bibfield  {title} {\bibinfo {title} {Relevance of the quadratic diamagnetic
  and self-polarization terms in cavity quantum electrodynamics},\ }\href
  {https://doi.org/10.1021/acsphotonics.9b01649} {\bibfield  {journal}
  {\bibinfo  {journal} {ACS Photonics}\ }\textbf {\bibinfo {volume} {7}},\
  \bibinfo {pages} {975} (\bibinfo {year} {2020})}\BibitemShut {NoStop}%
\bibitem [{\citenamefont {Ruggenthaler}\ \emph {et~al.}(2022)\citenamefont
  {Ruggenthaler}, \citenamefont {Sidler},\ and\ \citenamefont
  {Rubio}}]{Ruggenthaler2022}%
  \BibitemOpen
  \bibfield  {author} {\bibinfo {author} {\bibfnamefont {M.}~\bibnamefont
  {Ruggenthaler}}, \bibinfo {author} {\bibfnamefont {D.}~\bibnamefont
  {Sidler}},\ and\ \bibinfo {author} {\bibfnamefont {A.}~\bibnamefont
  {Rubio}},\ }\bibfield  {title} {\bibinfo {title} {Understanding polaritonic
  chemistry from ab initio quantum electrodynamics},\ }\bibfield  {journal}
  {\bibinfo  {journal} {arXiv:2211.04241}\ }\href
  {https://doi.org/10.48550/ARXIV.2211.04241} {10.48550/ARXIV.2211.04241}
  (\bibinfo {year} {2022})\BibitemShut {NoStop}%
\bibitem [{\citenamefont {Ruggenthaler}\ \emph {et~al.}(2014)\citenamefont
  {Ruggenthaler}, \citenamefont {Flick}, \citenamefont {Pellegrini},
  \citenamefont {Appel}, \citenamefont {Tokatly},\ and\ \citenamefont
  {Rubio}}]{Ruggenthaler2014}%
  \BibitemOpen
  \bibfield  {author} {\bibinfo {author} {\bibfnamefont {M.}~\bibnamefont
  {Ruggenthaler}}, \bibinfo {author} {\bibfnamefont {J.}~\bibnamefont {Flick}},
  \bibinfo {author} {\bibfnamefont {C.}~\bibnamefont {Pellegrini}}, \bibinfo
  {author} {\bibfnamefont {H.}~\bibnamefont {Appel}}, \bibinfo {author}
  {\bibfnamefont {I.~V.}\ \bibnamefont {Tokatly}},\ and\ \bibinfo {author}
  {\bibfnamefont {A.}~\bibnamefont {Rubio}},\ }\bibfield  {title} {\bibinfo
  {title} {Quantum-electrodynamical density-functional theory: Bridging quantum
  optics and electronic-structure theory},\ }\href
  {https://doi.org/10.1103/PhysRevA.90.012508} {\bibfield  {journal} {\bibinfo
  {journal} {Phys. Rev. A}\ }\textbf {\bibinfo {volume} {90}},\ \bibinfo
  {pages} {012508} (\bibinfo {year} {2014})}\BibitemShut {NoStop}%
\bibitem [{\citenamefont {Flick}\ \emph {et~al.}(2017)\citenamefont {Flick},
  \citenamefont {Ruggenthaler}, \citenamefont {Appel},\ and\ \citenamefont
  {Rubio}}]{Flick2017}%
  \BibitemOpen
  \bibfield  {author} {\bibinfo {author} {\bibfnamefont {J.}~\bibnamefont
  {Flick}}, \bibinfo {author} {\bibfnamefont {M.}~\bibnamefont {Ruggenthaler}},
  \bibinfo {author} {\bibfnamefont {H.}~\bibnamefont {Appel}},\ and\ \bibinfo
  {author} {\bibfnamefont {A.}~\bibnamefont {Rubio}},\ }\bibfield  {title}
  {\bibinfo {title} {Atoms and molecules in cavities, from weak to strong
  coupling in quantum-electrodynamics ({QED}) chemistry},\ }\href
  {https://doi.org/10.1073/pnas.1615509114} {\bibfield  {journal} {\bibinfo
  {journal} {Proc. Natl. Acad. Sci. U.S.A.}\ }\textbf {\bibinfo {volume}
  {114}},\ \bibinfo {pages} {3026} (\bibinfo {year} {2017})}\BibitemShut
  {NoStop}%
\bibitem [{\citenamefont {Haugland}\ \emph {et~al.}(2020)\citenamefont
  {Haugland}, \citenamefont {Ronca}, \citenamefont {Kjønstad}, \citenamefont
  {Rubio},\ and\ \citenamefont {Koch}}]{haugland_coupled_2020}%
  \BibitemOpen
  \bibfield  {author} {\bibinfo {author} {\bibfnamefont {T.~S.}\ \bibnamefont
  {Haugland}}, \bibinfo {author} {\bibfnamefont {E.}~\bibnamefont {Ronca}},
  \bibinfo {author} {\bibfnamefont {E.~F.}\ \bibnamefont {Kjønstad}}, \bibinfo
  {author} {\bibfnamefont {A.}~\bibnamefont {Rubio}},\ and\ \bibinfo {author}
  {\bibfnamefont {H.}~\bibnamefont {Koch}},\ }\bibfield  {title} {\bibinfo
  {title} {Coupled cluster theory for molecular polaritons: Changing ground and
  excited states},\ }\href {https://doi.org/10.1103/PhysRevX.10.041043}
  {\bibfield  {journal} {\bibinfo  {journal} {Phys. Rev. X}\ }\textbf {\bibinfo
  {volume} {10}},\ \bibinfo {pages} {041043} (\bibinfo {year}
  {2020})}\BibitemShut {NoStop}%
\bibitem [{\citenamefont {Mordovina}\ \emph {et~al.}(2020)\citenamefont
  {Mordovina}, \citenamefont {Bungey}, \citenamefont {Appel}, \citenamefont
  {Knowles}, \citenamefont {Rubio},\ and\ \citenamefont
  {Manby}}]{mordovina_polaritonic_2020}%
  \BibitemOpen
  \bibfield  {author} {\bibinfo {author} {\bibfnamefont {U.}~\bibnamefont
  {Mordovina}}, \bibinfo {author} {\bibfnamefont {C.}~\bibnamefont {Bungey}},
  \bibinfo {author} {\bibfnamefont {H.}~\bibnamefont {Appel}}, \bibinfo
  {author} {\bibfnamefont {P.~J.}\ \bibnamefont {Knowles}}, \bibinfo {author}
  {\bibfnamefont {A.}~\bibnamefont {Rubio}},\ and\ \bibinfo {author}
  {\bibfnamefont {F.~R.}\ \bibnamefont {Manby}},\ }\bibfield  {title} {\bibinfo
  {title} {Polaritonic coupled-cluster theory},\ }\href
  {https://doi.org/10.1103/PhysRevResearch.2.023262} {\bibfield  {journal}
  {\bibinfo  {journal} {Phys. Rev. Res.}\ }\textbf {\bibinfo {volume} {2}},\
  \bibinfo {pages} {023262} (\bibinfo {year} {2020})}\BibitemShut {NoStop}%
\bibitem [{\citenamefont {Pavo{\v s}evi{\'c}}\ and\ \citenamefont
  {Flick}(2021)}]{pavosevic_polaritonic_2021}%
  \BibitemOpen
  \bibfield  {author} {\bibinfo {author} {\bibfnamefont {F.}~\bibnamefont
  {Pavo{\v s}evi{\'c}}}\ and\ \bibinfo {author} {\bibfnamefont
  {J.}~\bibnamefont {Flick}},\ }\bibfield  {title} {\bibinfo {title}
  {Polaritonic unitary coupled cluster for quantum computations},\ }\href
  {https://doi.org/10.1021/acs.jpclett.1c02659} {\bibfield  {journal} {\bibinfo
   {journal} {J. Phys. Chem. Lett.}\ }\textbf {\bibinfo {volume} {12}},\
  \bibinfo {pages} {9100} (\bibinfo {year} {2021})}\BibitemShut {NoStop}%
\bibitem [{\citenamefont {Rohn}\ \emph {et~al.}(2020)\citenamefont {Rohn},
  \citenamefont {H\"ormann}, \citenamefont {Genes},\ and\ \citenamefont
  {Schmidt}}]{Rohn2020}%
  \BibitemOpen
  \bibfield  {author} {\bibinfo {author} {\bibfnamefont {J.}~\bibnamefont
  {Rohn}}, \bibinfo {author} {\bibfnamefont {M.}~\bibnamefont {H\"ormann}},
  \bibinfo {author} {\bibfnamefont {C.}~\bibnamefont {Genes}},\ and\ \bibinfo
  {author} {\bibfnamefont {K.~P.}\ \bibnamefont {Schmidt}},\ }\bibfield
  {title} {\bibinfo {title} {Ising model in a light-induced quantized
  transverse field},\ }\href {https://doi.org/10.1103/PhysRevResearch.2.023131}
  {\bibfield  {journal} {\bibinfo  {journal} {Phys. Rev. Res.}\ }\textbf
  {\bibinfo {volume} {2}},\ \bibinfo {pages} {023131} (\bibinfo {year}
  {2020})}\BibitemShut {NoStop}%
\bibitem [{\citenamefont {Schuler}\ \emph {et~al.}(2020)\citenamefont
  {Schuler}, \citenamefont {Bernardis}, \citenamefont {Läuchli},\ and\
  \citenamefont {Rabl}}]{Schuler2020}%
  \BibitemOpen
  \bibfield  {author} {\bibinfo {author} {\bibfnamefont {M.}~\bibnamefont
  {Schuler}}, \bibinfo {author} {\bibfnamefont {D.~D.}\ \bibnamefont
  {Bernardis}}, \bibinfo {author} {\bibfnamefont {A.~M.}\ \bibnamefont
  {Läuchli}},\ and\ \bibinfo {author} {\bibfnamefont {P.}~\bibnamefont
  {Rabl}},\ }\bibfield  {title} {\bibinfo {title} {{The Vacua of Dipolar Cavity
  Quantum Electrodynamics}},\ }\href
  {https://doi.org/10.21468/SciPostPhys.9.5.066} {\bibfield  {journal}
  {\bibinfo  {journal} {SciPost Phys.}\ }\textbf {\bibinfo {volume} {9}},\
  \bibinfo {pages} {66} (\bibinfo {year} {2020})}\BibitemShut {NoStop}%
\bibitem [{\citenamefont {M\'endez-C\'ordoba}\ \emph
  {et~al.}(2020)\citenamefont {M\'endez-C\'ordoba}, \citenamefont
  {Mendoza-Arenas}, \citenamefont {G\'omez-Ruiz}, \citenamefont
  {Rodr\'{\i}guez}, \citenamefont {Tejedor},\ and\ \citenamefont
  {Quiroga}}]{MendezCordoba2020}%
  \BibitemOpen
  \bibfield  {author} {\bibinfo {author} {\bibfnamefont {F.~P.~M.}\
  \bibnamefont {M\'endez-C\'ordoba}}, \bibinfo {author} {\bibfnamefont {J.~J.}\
  \bibnamefont {Mendoza-Arenas}}, \bibinfo {author} {\bibfnamefont {F.~J.}\
  \bibnamefont {G\'omez-Ruiz}}, \bibinfo {author} {\bibfnamefont {F.~J.}\
  \bibnamefont {Rodr\'{\i}guez}}, \bibinfo {author} {\bibfnamefont
  {C.}~\bibnamefont {Tejedor}},\ and\ \bibinfo {author} {\bibfnamefont
  {L.}~\bibnamefont {Quiroga}},\ }\bibfield  {title} {\bibinfo {title} {R\'enyi
  entropy singularities as signatures of topological criticality in coupled
  photon-fermion systems},\ }\href
  {https://doi.org/10.1103/PhysRevResearch.2.043264} {\bibfield  {journal}
  {\bibinfo  {journal} {Phys. Rev. Res.}\ }\textbf {\bibinfo {volume} {2}},\
  \bibinfo {pages} {043264} (\bibinfo {year} {2020})}\BibitemShut {NoStop}%
\bibitem [{\citenamefont {Chiriac\`o}\ \emph {et~al.}(2022)\citenamefont
  {Chiriac\`o}, \citenamefont {Dalmonte},\ and\ \citenamefont
  {Chanda}}]{Chiriaco2022}%
  \BibitemOpen
  \bibfield  {author} {\bibinfo {author} {\bibfnamefont {G.}~\bibnamefont
  {Chiriac\`o}}, \bibinfo {author} {\bibfnamefont {M.}~\bibnamefont
  {Dalmonte}},\ and\ \bibinfo {author} {\bibfnamefont {T.}~\bibnamefont
  {Chanda}},\ }\bibfield  {title} {\bibinfo {title} {Critical light-matter
  entanglement at cavity mediated phase transitions},\ }\href
  {https://doi.org/10.1103/PhysRevB.106.155113} {\bibfield  {journal} {\bibinfo
   {journal} {Phys. Rev. B}\ }\textbf {\bibinfo {volume} {106}},\ \bibinfo
  {pages} {155113} (\bibinfo {year} {2022})}\BibitemShut {NoStop}%
\bibitem [{\citenamefont {Passetti}\ \emph {et~al.}(2022)\citenamefont
  {Passetti}, \citenamefont {Eckhardt}, \citenamefont {Sentef},\ and\
  \citenamefont {Kennes}}]{Passetti2022}%
  \BibitemOpen
  \bibfield  {author} {\bibinfo {author} {\bibfnamefont {G.}~\bibnamefont
  {Passetti}}, \bibinfo {author} {\bibfnamefont {C.~J.}\ \bibnamefont
  {Eckhardt}}, \bibinfo {author} {\bibfnamefont {M.~A.}\ \bibnamefont
  {Sentef}},\ and\ \bibinfo {author} {\bibfnamefont {D.~M.}\ \bibnamefont
  {Kennes}},\ }\bibfield  {title} {\bibinfo {title} {Cavity light-matter
  entanglement through quantum fluctuations},\ }\bibfield  {journal} {\bibinfo
  {journal} {arXiv:2212.03011}\ }\href
  {https://doi.org/10.48550/ARXIV.2212.03011} {10.48550/ARXIV.2212.03011}
  (\bibinfo {year} {2022})\BibitemShut {NoStop}%
\bibitem [{\citenamefont {Novoselov}\ \emph {et~al.}(2016)\citenamefont
  {Novoselov}, \citenamefont {Mishchenko}, \citenamefont {Carvalho},\ and\
  \citenamefont {Castro~Neto}}]{novoselov_2d_2016}%
  \BibitemOpen
  \bibfield  {author} {\bibinfo {author} {\bibfnamefont {K.~S.}\ \bibnamefont
  {Novoselov}}, \bibinfo {author} {\bibfnamefont {A.}~\bibnamefont
  {Mishchenko}}, \bibinfo {author} {\bibfnamefont {A.}~\bibnamefont
  {Carvalho}},\ and\ \bibinfo {author} {\bibfnamefont {A.~H.}\ \bibnamefont
  {Castro~Neto}},\ }\bibfield  {title} {\bibinfo {title} {{{2D}} materials and
  van der {{Waals}} heterostructures},\ }\href
  {https://doi.org/10.1126/science.aac9439} {\bibfield  {journal} {\bibinfo
  {journal} {Science}\ }\textbf {\bibinfo {volume} {353}},\ \bibinfo {pages}
  {aac9439} (\bibinfo {year} {2016})}\BibitemShut {NoStop}%
\bibitem [{\citenamefont {Wang}\ \emph {et~al.}(2022)\citenamefont {Wang},
  \citenamefont {Bedoya-Pinto}, \citenamefont {Blei}, \citenamefont {Dismukes},
  \citenamefont {Hamo}, \citenamefont {Jenkins}, \citenamefont {Koperski},
  \citenamefont {Liu}, \citenamefont {Sun}, \citenamefont {Telford},
  \citenamefont {Kim}, \citenamefont {Augustin}, \citenamefont {Vool},
  \citenamefont {Yin}, \citenamefont {Li}, \citenamefont {Falin}, \citenamefont
  {Dean}, \citenamefont {Casanova}, \citenamefont {Evans}, \citenamefont
  {Chshiev}, \citenamefont {Mishchenko}, \citenamefont {Petrovic},
  \citenamefont {He}, \citenamefont {Zhao}, \citenamefont {Tsen}, \citenamefont
  {Gerardot}, \citenamefont {Brotons-Gisbert}, \citenamefont {Guguchia},
  \citenamefont {Roy}, \citenamefont {Tongay}, \citenamefont {Wang},
  \citenamefont {Hasan}, \citenamefont {Wrachtrup}, \citenamefont {Yacoby},
  \citenamefont {Fert}, \citenamefont {Parkin}, \citenamefont {Novoselov},
  \citenamefont {Dai}, \citenamefont {Balicas},\ and\ \citenamefont
  {Santos}}]{wang_magnetic_2022}%
  \BibitemOpen
  \bibfield  {author} {\bibinfo {author} {\bibfnamefont {Q.~H.}\ \bibnamefont
  {Wang}}, \bibinfo {author} {\bibfnamefont {A.}~\bibnamefont {Bedoya-Pinto}},
  \bibinfo {author} {\bibfnamefont {M.}~\bibnamefont {Blei}}, \bibinfo {author}
  {\bibfnamefont {A.~H.}\ \bibnamefont {Dismukes}}, \bibinfo {author}
  {\bibfnamefont {A.}~\bibnamefont {Hamo}}, \bibinfo {author} {\bibfnamefont
  {S.}~\bibnamefont {Jenkins}}, \bibinfo {author} {\bibfnamefont
  {M.}~\bibnamefont {Koperski}}, \bibinfo {author} {\bibfnamefont
  {Y.}~\bibnamefont {Liu}}, \bibinfo {author} {\bibfnamefont {Q.-C.}\
  \bibnamefont {Sun}}, \bibinfo {author} {\bibfnamefont {E.~J.}\ \bibnamefont
  {Telford}}, \bibinfo {author} {\bibfnamefont {H.~H.}\ \bibnamefont {Kim}},
  \bibinfo {author} {\bibfnamefont {M.}~\bibnamefont {Augustin}}, \bibinfo
  {author} {\bibfnamefont {U.}~\bibnamefont {Vool}}, \bibinfo {author}
  {\bibfnamefont {J.-X.}\ \bibnamefont {Yin}}, \bibinfo {author} {\bibfnamefont
  {L.~H.}\ \bibnamefont {Li}}, \bibinfo {author} {\bibfnamefont
  {A.}~\bibnamefont {Falin}}, \bibinfo {author} {\bibfnamefont {C.~R.}\
  \bibnamefont {Dean}}, \bibinfo {author} {\bibfnamefont {F.}~\bibnamefont
  {Casanova}}, \bibinfo {author} {\bibfnamefont {R.~F.~L.}\ \bibnamefont
  {Evans}}, \bibinfo {author} {\bibfnamefont {M.}~\bibnamefont {Chshiev}},
  \bibinfo {author} {\bibfnamefont {A.}~\bibnamefont {Mishchenko}}, \bibinfo
  {author} {\bibfnamefont {C.}~\bibnamefont {Petrovic}}, \bibinfo {author}
  {\bibfnamefont {R.}~\bibnamefont {He}}, \bibinfo {author} {\bibfnamefont
  {L.}~\bibnamefont {Zhao}}, \bibinfo {author} {\bibfnamefont {A.~W.}\
  \bibnamefont {Tsen}}, \bibinfo {author} {\bibfnamefont {B.~D.}\ \bibnamefont
  {Gerardot}}, \bibinfo {author} {\bibfnamefont {M.}~\bibnamefont
  {Brotons-Gisbert}}, \bibinfo {author} {\bibfnamefont {Z.}~\bibnamefont
  {Guguchia}}, \bibinfo {author} {\bibfnamefont {X.}~\bibnamefont {Roy}},
  \bibinfo {author} {\bibfnamefont {S.}~\bibnamefont {Tongay}}, \bibinfo
  {author} {\bibfnamefont {Z.}~\bibnamefont {Wang}}, \bibinfo {author}
  {\bibfnamefont {M.~Z.}\ \bibnamefont {Hasan}}, \bibinfo {author}
  {\bibfnamefont {J.}~\bibnamefont {Wrachtrup}}, \bibinfo {author}
  {\bibfnamefont {A.}~\bibnamefont {Yacoby}}, \bibinfo {author} {\bibfnamefont
  {A.}~\bibnamefont {Fert}}, \bibinfo {author} {\bibfnamefont {S.}~\bibnamefont
  {Parkin}}, \bibinfo {author} {\bibfnamefont {K.~S.}\ \bibnamefont
  {Novoselov}}, \bibinfo {author} {\bibfnamefont {P.}~\bibnamefont {Dai}},
  \bibinfo {author} {\bibfnamefont {L.}~\bibnamefont {Balicas}},\ and\ \bibinfo
  {author} {\bibfnamefont {E.~J.~G.}\ \bibnamefont {Santos}},\ }\bibfield
  {title} {\bibinfo {title} {The magnetic genome of two-dimensional van der
  {Waals} materials},\ }\href {https://doi.org/10.1021/acsnano.1c09150}
  {\bibfield  {journal} {\bibinfo  {journal} {ACS Nano}\ }\textbf {\bibinfo
  {volume} {16}},\ \bibinfo {pages} {6960} (\bibinfo {year}
  {2022})}\BibitemShut {NoStop}%
\bibitem [{\citenamefont {Rahman}\ \emph {et~al.}(2021)\citenamefont {Rahman},
  \citenamefont {Torres}, \citenamefont {Khan},\ and\ \citenamefont
  {Lu}}]{rahman_recent_2021}%
  \BibitemOpen
  \bibfield  {author} {\bibinfo {author} {\bibfnamefont {S.}~\bibnamefont
  {Rahman}}, \bibinfo {author} {\bibfnamefont {J.~F.}\ \bibnamefont {Torres}},
  \bibinfo {author} {\bibfnamefont {A.~R.}\ \bibnamefont {Khan}},\ and\
  \bibinfo {author} {\bibfnamefont {Y.}~\bibnamefont {Lu}},\ }\bibfield
  {title} {\bibinfo {title} {Recent developments in van der {Waals}
  antiferromagnetic {2D} materials: Synthesis, characterization, and device
  implementation},\ }\href {https://doi.org/10.1021/acsnano.1c06864} {\bibfield
   {journal} {\bibinfo  {journal} {ACS Nano}\ }\textbf {\bibinfo {volume}
  {15}},\ \bibinfo {pages} {17175} (\bibinfo {year} {2021})}\BibitemShut
  {NoStop}%
\bibitem [{\citenamefont {Wildes}\ \emph {et~al.}(1998)\citenamefont {Wildes},
  \citenamefont {Roessli}, \citenamefont {Lebech},\ and\ \citenamefont
  {Godfrey}}]{wildes_spin_1998}%
  \BibitemOpen
  \bibfield  {author} {\bibinfo {author} {\bibfnamefont {A.~R.}\ \bibnamefont
  {Wildes}}, \bibinfo {author} {\bibfnamefont {B.}~\bibnamefont {Roessli}},
  \bibinfo {author} {\bibfnamefont {B.}~\bibnamefont {Lebech}},\ and\ \bibinfo
  {author} {\bibfnamefont {K.~W.}\ \bibnamefont {Godfrey}},\ }\bibfield
  {title} {\bibinfo {title} {Spin waves and the critical behaviour of the
  magnetization in {MnPS$_3$}},\ }\href
  {https://doi.org/10.1088/0953-8984/10/28/020} {\bibfield  {journal} {\bibinfo
   {journal} {J. Phys. Condens. Matter}\ }\textbf {\bibinfo {volume} {10}},\
  \bibinfo {pages} {6417} (\bibinfo {year} {1998})}\BibitemShut {NoStop}%
\bibitem [{\citenamefont {Chittari}\ \emph {et~al.}(2016)\citenamefont
  {Chittari}, \citenamefont {Park}, \citenamefont {Lee}, \citenamefont {Han},
  \citenamefont {MacDonald}, \citenamefont {Hwang},\ and\ \citenamefont
  {Jung}}]{chittari_electronic_2016}%
  \BibitemOpen
  \bibfield  {author} {\bibinfo {author} {\bibfnamefont {B.~L.}\ \bibnamefont
  {Chittari}}, \bibinfo {author} {\bibfnamefont {Y.}~\bibnamefont {Park}},
  \bibinfo {author} {\bibfnamefont {D.}~\bibnamefont {Lee}}, \bibinfo {author}
  {\bibfnamefont {M.}~\bibnamefont {Han}}, \bibinfo {author} {\bibfnamefont
  {A.~H.}\ \bibnamefont {MacDonald}}, \bibinfo {author} {\bibfnamefont
  {E.}~\bibnamefont {Hwang}},\ and\ \bibinfo {author} {\bibfnamefont
  {J.}~\bibnamefont {Jung}},\ }\bibfield  {title} {\bibinfo {title} {Electronic
  and magnetic properties of single-layer {$M\mathrm{P}{X}_{3}$} metal
  phosphorous trichalcogenides},\ }\href
  {https://doi.org/10.1103/PhysRevB.94.184428} {\bibfield  {journal} {\bibinfo
  {journal} {Phys. Rev. B}\ }\textbf {\bibinfo {volume} {94}},\ \bibinfo
  {pages} {184428} (\bibinfo {year} {2016})}\BibitemShut {NoStop}%
\bibitem [{\citenamefont {Calder}\ \emph {et~al.}(2021)\citenamefont {Calder},
  \citenamefont {Haglund}, \citenamefont {Kolesnikov},\ and\ \citenamefont
  {Mandrus}}]{calder_magnetic_2021}%
  \BibitemOpen
  \bibfield  {author} {\bibinfo {author} {\bibfnamefont {S.}~\bibnamefont
  {Calder}}, \bibinfo {author} {\bibfnamefont {A.~V.}\ \bibnamefont {Haglund}},
  \bibinfo {author} {\bibfnamefont {A.~I.}\ \bibnamefont {Kolesnikov}},\ and\
  \bibinfo {author} {\bibfnamefont {D.}~\bibnamefont {Mandrus}},\ }\bibfield
  {title} {\bibinfo {title} {Magnetic exchange interactions in the van der
  {Waals} layered antiferromagnet {$\mathrm{Mn}\mathrm{P}{\mathrm{Se}}_{3}$}},\
  }\href {https://doi.org/10.1103/PhysRevB.103.024414} {\bibfield  {journal}
  {\bibinfo  {journal} {Phys. Rev. B}\ }\textbf {\bibinfo {volume} {103}},\
  \bibinfo {pages} {024414} (\bibinfo {year} {2021})}\BibitemShut {NoStop}%
\bibitem [{\citenamefont {Millis}\ and\ \citenamefont
  {Monien}(1993)}]{millis_spin_1993}%
  \BibitemOpen
  \bibfield  {author} {\bibinfo {author} {\bibfnamefont {A.~J.}\ \bibnamefont
  {Millis}}\ and\ \bibinfo {author} {\bibfnamefont {H.}~\bibnamefont
  {Monien}},\ }\bibfield  {title} {\bibinfo {title} {Spin gaps and spin
  dynamics in
  ${\mathrm{la}}_{2\ensuremath{-}x}{\mathrm{sr}}_{x}\mathrm{Cu}{\mathrm{o}}_{4}$
  and
  $\mathrm{Y}{\mathrm{ba}}_{2}{\mathrm{cu}}_{3}{\mathrm{o}}_{7\ensuremath{-}\ensuremath{\delta}}$},\
  }\href {https://doi.org/10.1103/PhysRevLett.70.2810} {\bibfield  {journal}
  {\bibinfo  {journal} {Phys. Rev. Lett.}\ }\textbf {\bibinfo {volume} {70}},\
  \bibinfo {pages} {2810} (\bibinfo {year} {1993})}\BibitemShut {NoStop}%
\bibitem [{\citenamefont {Chubukov}\ and\ \citenamefont
  {Morr}(1995)}]{chubukov_phase_1995}%
  \BibitemOpen
  \bibfield  {author} {\bibinfo {author} {\bibfnamefont {A.~V.}\ \bibnamefont
  {Chubukov}}\ and\ \bibinfo {author} {\bibfnamefont {D.~K.}\ \bibnamefont
  {Morr}},\ }\bibfield  {title} {\bibinfo {title} {Phase transition,
  longitudinal spin fluctuations, and scaling in a two-layer antiferromagnet},\
  }\href {https://doi.org/10.1103/PhysRevB.52.3521} {\bibfield  {journal}
  {\bibinfo  {journal} {Phys. Rev. B}\ }\textbf {\bibinfo {volume} {52}},\
  \bibinfo {pages} {3521} (\bibinfo {year} {1995})}\BibitemShut {NoStop}%
\bibitem [{\citenamefont {Sandvik}\ and\ \citenamefont
  {Scalapino}(1994)}]{sandvik_order-disorder_1994}%
  \BibitemOpen
  \bibfield  {author} {\bibinfo {author} {\bibfnamefont {A.~W.}\ \bibnamefont
  {Sandvik}}\ and\ \bibinfo {author} {\bibfnamefont {D.~J.}\ \bibnamefont
  {Scalapino}},\ }\bibfield  {title} {\bibinfo {title} {Order-disorder
  transition in a two-layer quantum antiferromagnet},\ }\href
  {https://doi.org/10.1103/PhysRevLett.72.2777} {\bibfield  {journal} {\bibinfo
   {journal} {Phys. Rev. Lett.}\ }\textbf {\bibinfo {volume} {72}},\ \bibinfo
  {pages} {2777} (\bibinfo {year} {1994})}\BibitemShut {NoStop}%
\bibitem [{\citenamefont {Ganesh}\ \emph {et~al.}(2011)\citenamefont {Ganesh},
  \citenamefont {Isakov},\ and\ \citenamefont {Paramekanti}}]{Ganesh2011}%
  \BibitemOpen
  \bibfield  {author} {\bibinfo {author} {\bibfnamefont {R.}~\bibnamefont
  {Ganesh}}, \bibinfo {author} {\bibfnamefont {S.~V.}\ \bibnamefont {Isakov}},\
  and\ \bibinfo {author} {\bibfnamefont {A.}~\bibnamefont {Paramekanti}},\
  }\bibfield  {title} {\bibinfo {title} {N\'eel to dimer transition in spin-$s$
  antiferromagnets: Comparing bond operator theory with quantum {M}onte {C}arlo
  simulations for bilayer {H}eisenberg models},\ }\href
  {https://doi.org/10.1103/PhysRevB.84.214412} {\bibfield  {journal} {\bibinfo
  {journal} {Phys. Rev. B}\ }\textbf {\bibinfo {volume} {84}},\ \bibinfo
  {pages} {214412} (\bibinfo {year} {2011})}\BibitemShut {NoStop}%
\bibitem [{\citenamefont {Song}\ \emph {et~al.}(2019)\citenamefont {Song},
  \citenamefont {Fei}, \citenamefont {Yankowitz}, \citenamefont {Lin},
  \citenamefont {Jiang}, \citenamefont {Hwangbo}, \citenamefont {Zhang},
  \citenamefont {Sun}, \citenamefont {Taniguchi}, \citenamefont {Watanabe},
  \citenamefont {McGuire}, \citenamefont {Graf}, \citenamefont {Cao},
  \citenamefont {Chu}, \citenamefont {Cobden}, \citenamefont {Dean},
  \citenamefont {Xiao},\ and\ \citenamefont {Xu}}]{song_switching_2019}%
  \BibitemOpen
  \bibfield  {author} {\bibinfo {author} {\bibfnamefont {T.}~\bibnamefont
  {Song}}, \bibinfo {author} {\bibfnamefont {Z.}~\bibnamefont {Fei}}, \bibinfo
  {author} {\bibfnamefont {M.}~\bibnamefont {Yankowitz}}, \bibinfo {author}
  {\bibfnamefont {Z.}~\bibnamefont {Lin}}, \bibinfo {author} {\bibfnamefont
  {Q.}~\bibnamefont {Jiang}}, \bibinfo {author} {\bibfnamefont
  {K.}~\bibnamefont {Hwangbo}}, \bibinfo {author} {\bibfnamefont
  {Q.}~\bibnamefont {Zhang}}, \bibinfo {author} {\bibfnamefont
  {B.}~\bibnamefont {Sun}}, \bibinfo {author} {\bibfnamefont {T.}~\bibnamefont
  {Taniguchi}}, \bibinfo {author} {\bibfnamefont {K.}~\bibnamefont {Watanabe}},
  \bibinfo {author} {\bibfnamefont {M.~A.}\ \bibnamefont {McGuire}}, \bibinfo
  {author} {\bibfnamefont {D.}~\bibnamefont {Graf}}, \bibinfo {author}
  {\bibfnamefont {T.}~\bibnamefont {Cao}}, \bibinfo {author} {\bibfnamefont
  {J.-H.}\ \bibnamefont {Chu}}, \bibinfo {author} {\bibfnamefont {D.~H.}\
  \bibnamefont {Cobden}}, \bibinfo {author} {\bibfnamefont {C.~R.}\
  \bibnamefont {Dean}}, \bibinfo {author} {\bibfnamefont {D.}~\bibnamefont
  {Xiao}},\ and\ \bibinfo {author} {\bibfnamefont {X.}~\bibnamefont {Xu}},\
  }\bibfield  {title} {\bibinfo {title} {Switching {{2D}} magnetic states via
  pressure tuning of layer stacking},\ }\href
  {https://doi.org/10.1038/s41563-019-0505-2} {\bibfield  {journal} {\bibinfo
  {journal} {Nat. Mater.}\ }\textbf {\bibinfo {volume} {18}},\ \bibinfo {pages}
  {1298} (\bibinfo {year} {2019})}\BibitemShut {NoStop}%
\bibitem [{\citenamefont {Saidl}\ \emph {et~al.}(2017)\citenamefont {Saidl},
  \citenamefont {N{\v e}mec}, \citenamefont {Wadley}, \citenamefont {Hills},
  \citenamefont {Campion}, \citenamefont {Nov{\'a}k}, \citenamefont {Edmonds},
  \citenamefont {Maccherozzi}, \citenamefont {Dhesi}, \citenamefont
  {Gallagher}, \citenamefont {Troj{\'a}nek}, \citenamefont {Kune{\v s}},
  \citenamefont {{\v Z}elezn{\'y}}, \citenamefont {Mal{\'y}},\ and\
  \citenamefont {Jungwirth}}]{saidl_optical_2017}%
  \BibitemOpen
  \bibfield  {author} {\bibinfo {author} {\bibfnamefont {V.}~\bibnamefont
  {Saidl}}, \bibinfo {author} {\bibfnamefont {P.}~\bibnamefont {N{\v e}mec}},
  \bibinfo {author} {\bibfnamefont {P.}~\bibnamefont {Wadley}}, \bibinfo
  {author} {\bibfnamefont {V.}~\bibnamefont {Hills}}, \bibinfo {author}
  {\bibfnamefont {R.~P.}\ \bibnamefont {Campion}}, \bibinfo {author}
  {\bibfnamefont {V.}~\bibnamefont {Nov{\'a}k}}, \bibinfo {author}
  {\bibfnamefont {K.~W.}\ \bibnamefont {Edmonds}}, \bibinfo {author}
  {\bibfnamefont {F.}~\bibnamefont {Maccherozzi}}, \bibinfo {author}
  {\bibfnamefont {S.~S.}\ \bibnamefont {Dhesi}}, \bibinfo {author}
  {\bibfnamefont {B.~L.}\ \bibnamefont {Gallagher}}, \bibinfo {author}
  {\bibfnamefont {F.}~\bibnamefont {Troj{\'a}nek}}, \bibinfo {author}
  {\bibfnamefont {J.}~\bibnamefont {Kune{\v s}}}, \bibinfo {author}
  {\bibfnamefont {J.}~\bibnamefont {{\v Z}elezn{\'y}}}, \bibinfo {author}
  {\bibfnamefont {P.}~\bibnamefont {Mal{\'y}}},\ and\ \bibinfo {author}
  {\bibfnamefont {T.}~\bibnamefont {Jungwirth}},\ }\bibfield  {title} {\bibinfo
  {title} {Optical determination of the {{N\'eel}} vector in a {{CuMnAs}}
  thin-film antiferromagnet},\ }\href
  {https://doi.org/10.1038/nphoton.2016.255} {\bibfield  {journal} {\bibinfo
  {journal} {Nat. Photonics}\ }\textbf {\bibinfo {volume} {11}},\ \bibinfo
  {pages} {91} (\bibinfo {year} {2017})}\BibitemShut {NoStop}%
\bibitem [{\citenamefont {Grigorev}\ \emph {et~al.}(2021)\citenamefont
  {Grigorev}, \citenamefont {Filianina}, \citenamefont {Bodnar}, \citenamefont
  {Sobolev}, \citenamefont {Bhattacharjee}, \citenamefont {Bommanaboyena},
  \citenamefont {Lytvynenko}, \citenamefont {Skourski}, \citenamefont {Fuchs},
  \citenamefont {Kl{\"a}ui}, \citenamefont {Jourdan},\ and\ \citenamefont
  {Demsar}}]{grigorev_optical_2021}%
  \BibitemOpen
  \bibfield  {author} {\bibinfo {author} {\bibfnamefont {V.}~\bibnamefont
  {Grigorev}}, \bibinfo {author} {\bibfnamefont {M.}~\bibnamefont {Filianina}},
  \bibinfo {author} {\bibfnamefont {S.~Y.}\ \bibnamefont {Bodnar}}, \bibinfo
  {author} {\bibfnamefont {S.}~\bibnamefont {Sobolev}}, \bibinfo {author}
  {\bibfnamefont {N.}~\bibnamefont {Bhattacharjee}}, \bibinfo {author}
  {\bibfnamefont {S.}~\bibnamefont {Bommanaboyena}}, \bibinfo {author}
  {\bibfnamefont {Y.}~\bibnamefont {Lytvynenko}}, \bibinfo {author}
  {\bibfnamefont {Y.}~\bibnamefont {Skourski}}, \bibinfo {author}
  {\bibfnamefont {D.}~\bibnamefont {Fuchs}}, \bibinfo {author} {\bibfnamefont
  {M.}~\bibnamefont {Kl{\"a}ui}}, \bibinfo {author} {\bibfnamefont
  {M.}~\bibnamefont {Jourdan}},\ and\ \bibinfo {author} {\bibfnamefont
  {J.}~\bibnamefont {Demsar}},\ }\bibfield  {title} {\bibinfo {title} {Optical
  readout of the {{N}}éel vector in the metallic antiferromagnet
  ${\mathrm{mn}}_{2}\mathrm{Au}$},\ }\href
  {https://doi.org/10.1103/PhysRevApplied.16.014037} {\bibfield  {journal}
  {\bibinfo  {journal} {Phys. Rev. Appl.}\ }\textbf {\bibinfo {volume} {16}},\
  \bibinfo {pages} {014037} (\bibinfo {year} {2021})}\BibitemShut {NoStop}%
\bibitem [{\citenamefont {Kim}\ \emph {et~al.}(2019)\citenamefont {Kim},
  \citenamefont {Lim}, \citenamefont {Kim}, \citenamefont {Lee}, \citenamefont
  {Lee}, \citenamefont {Kim}, \citenamefont {Park}, \citenamefont {Son},
  \citenamefont {Park}, \citenamefont {Park},\ and\ \citenamefont
  {Cheong}}]{kim_antiferromagnetic_2019-1}%
  \BibitemOpen
  \bibfield  {author} {\bibinfo {author} {\bibfnamefont {K.}~\bibnamefont
  {Kim}}, \bibinfo {author} {\bibfnamefont {S.~Y.}\ \bibnamefont {Lim}},
  \bibinfo {author} {\bibfnamefont {J.}~\bibnamefont {Kim}}, \bibinfo {author}
  {\bibfnamefont {J.-U.}\ \bibnamefont {Lee}}, \bibinfo {author} {\bibfnamefont
  {S.}~\bibnamefont {Lee}}, \bibinfo {author} {\bibfnamefont {P.}~\bibnamefont
  {Kim}}, \bibinfo {author} {\bibfnamefont {K.}~\bibnamefont {Park}}, \bibinfo
  {author} {\bibfnamefont {S.}~\bibnamefont {Son}}, \bibinfo {author}
  {\bibfnamefont {C.-H.}\ \bibnamefont {Park}}, \bibinfo {author}
  {\bibfnamefont {J.-G.}\ \bibnamefont {Park}},\ and\ \bibinfo {author}
  {\bibfnamefont {H.}~\bibnamefont {Cheong}},\ }\bibfield  {title} {\bibinfo
  {title} {Antiferromagnetic ordering in van der {{Waals 2D}} magnetic material
  {{MnPS\textsubscript{3}}} probed by {{Raman}} spectroscopy},\ }\href
  {https://doi.org/10.1088/2053-1583/ab27d5} {\bibfield  {journal} {\bibinfo
  {journal} {2D Mater.}\ }\textbf {\bibinfo {volume} {6}},\ \bibinfo {pages}
  {041001} (\bibinfo {year} {2019})}\BibitemShut {NoStop}%
\bibitem [{\citenamefont {Sandvik}\ \emph {et~al.}(1997)\citenamefont
  {Sandvik}, \citenamefont {Singh},\ and\ \citenamefont
  {Campbell}}]{Sandvik1997}%
  \BibitemOpen
  \bibfield  {author} {\bibinfo {author} {\bibfnamefont {A.~W.}\ \bibnamefont
  {Sandvik}}, \bibinfo {author} {\bibfnamefont {R.~R.~P.}\ \bibnamefont
  {Singh}},\ and\ \bibinfo {author} {\bibfnamefont {D.~K.}\ \bibnamefont
  {Campbell}},\ }\bibfield  {title} {\bibinfo {title} {Quantum {M}onte {C}arlo
  in the interaction representation: {A}pplication to a spin-{P}eierls model},\
  }\href {https://doi.org/10.1103/PhysRevB.56.14510} {\bibfield  {journal}
  {\bibinfo  {journal} {Phys. Rev. B}\ }\textbf {\bibinfo {volume} {56}},\
  \bibinfo {pages} {14510} (\bibinfo {year} {1997})}\BibitemShut {NoStop}%
\bibitem [{\citenamefont {Weber}(2022)}]{MWeber2022}%
  \BibitemOpen
  \bibfield  {author} {\bibinfo {author} {\bibfnamefont {M.}~\bibnamefont
  {Weber}},\ }\bibfield  {title} {\bibinfo {title} {Quantum {M}onte {C}arlo
  simulation of spin-boson models using wormhole updates},\ }\href
  {https://doi.org/10.1103/PhysRevB.105.165129} {\bibfield  {journal} {\bibinfo
   {journal} {Phys. Rev. B}\ }\textbf {\bibinfo {volume} {105}},\ \bibinfo
  {pages} {165129} (\bibinfo {year} {2022})}\BibitemShut {NoStop}%
\bibitem [{\citenamefont {Weber}\ \emph
  {et~al.}(2022{\natexlab{a}})\citenamefont {Weber}, \citenamefont {Luitz},\
  and\ \citenamefont {Assaad}}]{MWeber2022b}%
  \BibitemOpen
  \bibfield  {author} {\bibinfo {author} {\bibfnamefont {M.}~\bibnamefont
  {Weber}}, \bibinfo {author} {\bibfnamefont {D.~J.}\ \bibnamefont {Luitz}},\
  and\ \bibinfo {author} {\bibfnamefont {F.~F.}\ \bibnamefont {Assaad}},\
  }\bibfield  {title} {\bibinfo {title} {Dissipation-induced order: The
  {$S=1/2$} quantum spin chain coupled to an ohmic bath},\ }\href
  {https://doi.org/10.1103/PhysRevLett.129.056402} {\bibfield  {journal}
  {\bibinfo  {journal} {Phys. Rev. Lett.}\ }\textbf {\bibinfo {volume} {129}},\
  \bibinfo {pages} {056402} (\bibinfo {year} {2022}{\natexlab{a}})}\BibitemShut
  {NoStop}%
\bibitem [{\citenamefont {Dmytruk}\ and\ \citenamefont
  {Schir\'o}(2021)}]{Dmytruk2021}%
  \BibitemOpen
  \bibfield  {author} {\bibinfo {author} {\bibfnamefont {O.}~\bibnamefont
  {Dmytruk}}\ and\ \bibinfo {author} {\bibfnamefont {M.}~\bibnamefont
  {Schir\'o}},\ }\bibfield  {title} {\bibinfo {title} {Gauge fixing for
  strongly correlated electrons coupled to quantum light},\ }\href
  {https://doi.org/10.1103/PhysRevB.103.075131} {\bibfield  {journal} {\bibinfo
   {journal} {Phys. Rev. B}\ }\textbf {\bibinfo {volume} {103}},\ \bibinfo
  {pages} {075131} (\bibinfo {year} {2021})}\BibitemShut {NoStop}%
\bibitem [{\citenamefont {Eckhardt}\ \emph {et~al.}(2022)\citenamefont
  {Eckhardt}, \citenamefont {Passetti}, \citenamefont {Othman}, \citenamefont
  {Karrasch}, \citenamefont {Cavaliere}, \citenamefont {Sentef},\ and\
  \citenamefont {Kennes}}]{Eckhardt2022}%
  \BibitemOpen
  \bibfield  {author} {\bibinfo {author} {\bibfnamefont {C.~J.}\ \bibnamefont
  {Eckhardt}}, \bibinfo {author} {\bibfnamefont {G.}~\bibnamefont {Passetti}},
  \bibinfo {author} {\bibfnamefont {M.}~\bibnamefont {Othman}}, \bibinfo
  {author} {\bibfnamefont {C.}~\bibnamefont {Karrasch}}, \bibinfo {author}
  {\bibfnamefont {F.}~\bibnamefont {Cavaliere}}, \bibinfo {author}
  {\bibfnamefont {M.~A.}\ \bibnamefont {Sentef}},\ and\ \bibinfo {author}
  {\bibfnamefont {D.~M.}\ \bibnamefont {Kennes}},\ }\bibfield  {title}
  {\bibinfo {title} {Quantum {F}loquet engineering with an exactly solvable
  tight-binding chain in a cavity},\ }\href
  {https://doi.org/10.1038/s42005-022-00880-9} {\bibfield  {journal} {\bibinfo
  {journal} {Comm. Phys.}\ }\textbf {\bibinfo {volume} {5}},\ \bibinfo {pages}
  {122} (\bibinfo {year} {2022})}\BibitemShut {NoStop}%
\bibitem [{\citenamefont {Kiffner}\ \emph {et~al.}(2019)\citenamefont
  {Kiffner}, \citenamefont {Coulthard}, \citenamefont {Schlawin}, \citenamefont
  {Ardavan},\ and\ \citenamefont {Jaksch}}]{Kiffner2019}%
  \BibitemOpen
  \bibfield  {author} {\bibinfo {author} {\bibfnamefont {M.}~\bibnamefont
  {Kiffner}}, \bibinfo {author} {\bibfnamefont {J.}~\bibnamefont {Coulthard}},
  \bibinfo {author} {\bibfnamefont {F.}~\bibnamefont {Schlawin}}, \bibinfo
  {author} {\bibfnamefont {A.}~\bibnamefont {Ardavan}},\ and\ \bibinfo {author}
  {\bibfnamefont {D.}~\bibnamefont {Jaksch}},\ }\bibfield  {title} {\bibinfo
  {title} {{M}ott polaritons in cavity-coupled quantum materials},\ }\href
  {https://doi.org/10.1088/1367-2630/ab31c7} {\bibfield  {journal} {\bibinfo
  {journal} {New J. Phys.}\ }\textbf {\bibinfo {volume} {21}},\ \bibinfo
  {pages} {073066} (\bibinfo {year} {2019})}\BibitemShut {NoStop}%
\bibitem [{\citenamefont {Sandvik}(1999)}]{Sandvik1999}%
  \BibitemOpen
  \bibfield  {author} {\bibinfo {author} {\bibfnamefont {A.~W.}\ \bibnamefont
  {Sandvik}},\ }\bibfield  {title} {\bibinfo {title} {Stochastic series
  expansion method with operator-loop update},\ }\href
  {https://doi.org/10.1103/PhysRevB.59.R14157} {\bibfield  {journal} {\bibinfo
  {journal} {Phys. Rev. B}\ }\textbf {\bibinfo {volume} {59}},\ \bibinfo
  {pages} {R14157} (\bibinfo {year} {1999})}\BibitemShut {NoStop}%
\bibitem [{\citenamefont {Sylju\aa{}sen}\ and\ \citenamefont
  {Sandvik}(2002)}]{Syljuasen2002}%
  \BibitemOpen
  \bibfield  {author} {\bibinfo {author} {\bibfnamefont {O.~F.}\ \bibnamefont
  {Sylju\aa{}sen}}\ and\ \bibinfo {author} {\bibfnamefont {A.~W.}\ \bibnamefont
  {Sandvik}},\ }\bibfield  {title} {\bibinfo {title} {Quantum {M}onte {C}arlo
  with directed loops},\ }\href {https://doi.org/10.1103/PhysRevE.66.046701}
  {\bibfield  {journal} {\bibinfo  {journal} {Phys. Rev. E}\ }\textbf {\bibinfo
  {volume} {66}},\ \bibinfo {pages} {046701} (\bibinfo {year}
  {2002})}\BibitemShut {NoStop}%
\bibitem [{\citenamefont {Pan}\ and\ \citenamefont {Meng}(2022)}]{Pan2022}%
  \BibitemOpen
  \bibfield  {author} {\bibinfo {author} {\bibfnamefont {G.}~\bibnamefont
  {Pan}}\ and\ \bibinfo {author} {\bibfnamefont {Z.~Y.}\ \bibnamefont {Meng}},\
  }\href@noop {} {\bibinfo {title} {Sign problem in quantum {M}onte {C}arlo
  simulation}} (\bibinfo {year} {2022}),\ \Eprint
  {https://arxiv.org/abs/arXix:2204.08777} {arXix:2204.08777} \BibitemShut
  {NoStop}%
\bibitem [{\citenamefont {Weber}\ \emph
  {et~al.}(2022{\natexlab{b}})\citenamefont {Weber}, \citenamefont {Honecker},
  \citenamefont {Normand}, \citenamefont {Corboz}, \citenamefont {Mila},\ and\
  \citenamefont {Wessel}}]{Weber2021}%
  \BibitemOpen
  \bibfield  {author} {\bibinfo {author} {\bibfnamefont {L.}~\bibnamefont
  {Weber}}, \bibinfo {author} {\bibfnamefont {A.}~\bibnamefont {Honecker}},
  \bibinfo {author} {\bibfnamefont {B.}~\bibnamefont {Normand}}, \bibinfo
  {author} {\bibfnamefont {P.}~\bibnamefont {Corboz}}, \bibinfo {author}
  {\bibfnamefont {F.}~\bibnamefont {Mila}},\ and\ \bibinfo {author}
  {\bibfnamefont {S.}~\bibnamefont {Wessel}},\ }\bibfield  {title} {\bibinfo
  {title} {Quantum {M}onte {C}arlo simulations in the trimer basis: first-order
  transitions and thermal critical points in frustrated trilayer magnets},\
  }\href {https://doi.org/10.21468/SciPostPhys.12.2.054} {\bibfield  {journal}
  {\bibinfo  {journal} {SciPost Phys.}\ }\textbf {\bibinfo {volume} {12}},\
  \bibinfo {pages} {54} (\bibinfo {year} {2022}{\natexlab{b}})}\BibitemShut
  {NoStop}%
\bibitem [{\citenamefont {Alet}\ and\ \citenamefont
  {S\o{}rensen}(2003)}]{Alet2003}%
  \BibitemOpen
  \bibfield  {author} {\bibinfo {author} {\bibfnamefont {F.}~\bibnamefont
  {Alet}}\ and\ \bibinfo {author} {\bibfnamefont {E.~S.}\ \bibnamefont
  {S\o{}rensen}},\ }\bibfield  {title} {\bibinfo {title} {Directed geometrical
  worm algorithm applied to the quantum rotor model},\ }\href
  {https://doi.org/10.1103/PhysRevE.68.026702} {\bibfield  {journal} {\bibinfo
  {journal} {Phys. Rev. E}\ }\textbf {\bibinfo {volume} {68}},\ \bibinfo
  {pages} {026702} (\bibinfo {year} {2003})}\BibitemShut {NoStop}%
\bibitem [{\citenamefont {Alet}\ \emph {et~al.}(2005)\citenamefont {Alet},
  \citenamefont {Wessel},\ and\ \citenamefont {Troyer}}]{Alet2005}%
  \BibitemOpen
  \bibfield  {author} {\bibinfo {author} {\bibfnamefont {F.}~\bibnamefont
  {Alet}}, \bibinfo {author} {\bibfnamefont {S.}~\bibnamefont {Wessel}},\ and\
  \bibinfo {author} {\bibfnamefont {M.}~\bibnamefont {Troyer}},\ }\bibfield
  {title} {\bibinfo {title} {Generalized directed loop method for quantum
  {M}onte {C}arlo simulations},\ }\href
  {https://doi.org/10.1103/PhysRevE.71.036706} {\bibfield  {journal} {\bibinfo
  {journal} {Phys. Rev. E}\ }\textbf {\bibinfo {volume} {71}},\ \bibinfo
  {pages} {036706} (\bibinfo {year} {2005})}\BibitemShut {NoStop}%
\bibitem [{\citenamefont {Huangfu}\ and\ \citenamefont
  {Hall}(2018)}]{Huangfu2018}%
  \BibitemOpen
  \bibfield  {author} {\bibinfo {author} {\bibfnamefont {Q.}~\bibnamefont
  {Huangfu}}\ and\ \bibinfo {author} {\bibfnamefont {J.~A.~J.}\ \bibnamefont
  {Hall}},\ }\bibfield  {title} {\bibinfo {title} {Parallelizing the dual
  revised simplex method},\ }\href {https://doi.org/10.1007/s12532-017-0130-5}
  {\bibfield  {journal} {\bibinfo  {journal} {Math. Program. Comput.}\ }\textbf
  {\bibinfo {volume} {10}},\ \bibinfo {pages} {119} (\bibinfo {year}
  {2018})}\BibitemShut {NoStop}%
\bibitem [{\citenamefont {Hardikar}\ and\ \citenamefont
  {Clay}(2007)}]{Hardikar2007}%
  \BibitemOpen
  \bibfield  {author} {\bibinfo {author} {\bibfnamefont {R.~P.}\ \bibnamefont
  {Hardikar}}\ and\ \bibinfo {author} {\bibfnamefont {R.~T.}\ \bibnamefont
  {Clay}},\ }\bibfield  {title} {\bibinfo {title} {Phase diagram of the
  one-dimensional {H}ubbard-{H}olstein model at half and quarter filling},\
  }\href {https://doi.org/10.1103/PhysRevB.75.245103} {\bibfield  {journal}
  {\bibinfo  {journal} {Phys. Rev. B}\ }\textbf {\bibinfo {volume} {75}},\
  \bibinfo {pages} {245103} (\bibinfo {year} {2007})}\BibitemShut {NoStop}%
\bibitem [{\citenamefont {Matsumoto}\ \emph {et~al.}(2001)\citenamefont
  {Matsumoto}, \citenamefont {Yasuda}, \citenamefont {Todo},\ and\
  \citenamefont {Takayama}}]{Matsumoto2001}%
  \BibitemOpen
  \bibfield  {author} {\bibinfo {author} {\bibfnamefont {M.}~\bibnamefont
  {Matsumoto}}, \bibinfo {author} {\bibfnamefont {C.}~\bibnamefont {Yasuda}},
  \bibinfo {author} {\bibfnamefont {S.}~\bibnamefont {Todo}},\ and\ \bibinfo
  {author} {\bibfnamefont {H.}~\bibnamefont {Takayama}},\ }\bibfield  {title}
  {\bibinfo {title} {Ground-state phase diagram of quantum {H}eisenberg
  antiferromagnets on the anisotropic dimerized square lattice},\ }\href
  {https://doi.org/10.1103/PhysRevB.65.014407} {\bibfield  {journal} {\bibinfo
  {journal} {Phys. Rev. B}\ }\textbf {\bibinfo {volume} {65}},\ \bibinfo
  {pages} {014407} (\bibinfo {year} {2001})}\BibitemShut {NoStop}%
\bibitem [{\citenamefont {Andolina}\ \emph {et~al.}(2019)\citenamefont
  {Andolina}, \citenamefont {Pellegrino}, \citenamefont {Giovannetti},
  \citenamefont {MacDonald},\ and\ \citenamefont {Polini}}]{Andolina2019}%
  \BibitemOpen
  \bibfield  {author} {\bibinfo {author} {\bibfnamefont {G.~M.}\ \bibnamefont
  {Andolina}}, \bibinfo {author} {\bibfnamefont {F.~M.~D.}\ \bibnamefont
  {Pellegrino}}, \bibinfo {author} {\bibfnamefont {V.}~\bibnamefont
  {Giovannetti}}, \bibinfo {author} {\bibfnamefont {A.~H.}\ \bibnamefont
  {MacDonald}},\ and\ \bibinfo {author} {\bibfnamefont {M.}~\bibnamefont
  {Polini}},\ }\bibfield  {title} {\bibinfo {title} {Cavity quantum
  electrodynamics of strongly correlated electron systems: A no-go theorem for
  photon condensation},\ }\href {https://doi.org/10.1103/PhysRevB.100.121109}
  {\bibfield  {journal} {\bibinfo  {journal} {Phys. Rev. B}\ }\textbf {\bibinfo
  {volume} {100}},\ \bibinfo {pages} {121109} (\bibinfo {year}
  {2019})}\BibitemShut {NoStop}%
\bibitem [{\citenamefont {Andolina}\ \emph {et~al.}(2020)\citenamefont
  {Andolina}, \citenamefont {Pellegrino}, \citenamefont {Giovannetti},
  \citenamefont {MacDonald},\ and\ \citenamefont {Polini}}]{Andolina2020}%
  \BibitemOpen
  \bibfield  {author} {\bibinfo {author} {\bibfnamefont {G.~M.}\ \bibnamefont
  {Andolina}}, \bibinfo {author} {\bibfnamefont {F.~M.~D.}\ \bibnamefont
  {Pellegrino}}, \bibinfo {author} {\bibfnamefont {V.}~\bibnamefont
  {Giovannetti}}, \bibinfo {author} {\bibfnamefont {A.~H.}\ \bibnamefont
  {MacDonald}},\ and\ \bibinfo {author} {\bibfnamefont {M.}~\bibnamefont
  {Polini}},\ }\bibfield  {title} {\bibinfo {title} {Theory of photon
  condensation in a spatially varying electromagnetic field},\ }\href
  {https://doi.org/10.1103/PhysRevB.102.125137} {\bibfield  {journal} {\bibinfo
   {journal} {Phys. Rev. B}\ }\textbf {\bibinfo {volume} {102}},\ \bibinfo
  {pages} {125137} (\bibinfo {year} {2020})}\BibitemShut {NoStop}%
\bibitem [{\citenamefont {Lenk}\ \emph {et~al.}(2022)\citenamefont {Lenk},
  \citenamefont {Li}, \citenamefont {Werner},\ and\ \citenamefont
  {Eckstein}}]{Lenk2022}%
  \BibitemOpen
  \bibfield  {author} {\bibinfo {author} {\bibfnamefont {K.}~\bibnamefont
  {Lenk}}, \bibinfo {author} {\bibfnamefont {J.}~\bibnamefont {Li}}, \bibinfo
  {author} {\bibfnamefont {P.}~\bibnamefont {Werner}},\ and\ \bibinfo {author}
  {\bibfnamefont {M.}~\bibnamefont {Eckstein}},\ }\href@noop {} {\bibinfo
  {title} {Collective theory for an interacting solid in a single-mode cavity}}
  (\bibinfo {year} {2022}),\ \Eprint {https://arxiv.org/abs/2205.05559}
  {arXiv:2205.05559} \BibitemShut {NoStop}%
\bibitem [{\citenamefont {Wegner}(1972)}]{wegner_corrections_1972}%
  \BibitemOpen
  \bibfield  {author} {\bibinfo {author} {\bibfnamefont {F.~J.}\ \bibnamefont
  {Wegner}},\ }\bibfield  {title} {\bibinfo {title} {Corrections to scaling
  laws},\ }\href {https://doi.org/10.1103/PhysRevB.5.4529} {\bibfield
  {journal} {\bibinfo  {journal} {Phys. Rev. B}\ }\textbf {\bibinfo {volume}
  {5}},\ \bibinfo {pages} {4529} (\bibinfo {year} {1972})}\BibitemShut
  {NoStop}%
\bibitem [{\citenamefont {Fradkin}\ and\ \citenamefont
  {Moore}(2006)}]{fradkin_entanglement_2006}%
  \BibitemOpen
  \bibfield  {author} {\bibinfo {author} {\bibfnamefont {E.}~\bibnamefont
  {Fradkin}}\ and\ \bibinfo {author} {\bibfnamefont {J.~E.}\ \bibnamefont
  {Moore}},\ }\bibfield  {title} {\bibinfo {title} {Entanglement entropy of
  {2D} conformal quantum critical points: Hearing the shape of a quantum
  drum},\ }\href {https://doi.org/10.1103/PhysRevLett.97.050404} {\bibfield
  {journal} {\bibinfo  {journal} {Phys. Rev. Lett.}\ }\textbf {\bibinfo
  {volume} {97}},\ \bibinfo {pages} {050404} (\bibinfo {year}
  {2006})}\BibitemShut {NoStop}%
\bibitem [{\citenamefont {Kos}\ \emph {et~al.}(2016)\citenamefont {Kos},
  \citenamefont {Poland}, \citenamefont {Simmons-Duffin},\ and\ \citenamefont
  {Vichi}}]{Kos2016}%
  \BibitemOpen
  \bibfield  {author} {\bibinfo {author} {\bibfnamefont {F.}~\bibnamefont
  {Kos}}, \bibinfo {author} {\bibfnamefont {D.}~\bibnamefont {Poland}},
  \bibinfo {author} {\bibfnamefont {D.}~\bibnamefont {Simmons-Duffin}},\ and\
  \bibinfo {author} {\bibfnamefont {A.}~\bibnamefont {Vichi}},\ }\bibfield
  {title} {\bibinfo {title} {Precision islands in the {I}sing and {O(N)}
  models},\ }\href {https://doi.org/10.1007/JHEP08(2016)036} {\bibfield
  {journal} {\bibinfo  {journal} {J. High Energy Phys.}\ }\textbf {\bibinfo
  {volume} {2016}}\bibinfo  {number} { (8)},\ \bibinfo {pages}
  {36}}\BibitemShut {NoStop}%
\bibitem [{\citenamefont {Sachdev}\ and\ \citenamefont
  {Bhatt}(1990)}]{Sachdev1990}%
  \BibitemOpen
\bibfield  {number} {  }\bibfield  {author} {\bibinfo {author} {\bibfnamefont
  {S.}~\bibnamefont {Sachdev}}\ and\ \bibinfo {author} {\bibfnamefont {R.~N.}\
  \bibnamefont {Bhatt}},\ }\bibfield  {title} {\bibinfo {title} {Bond-operator
  representation of quantum spins: Mean-field theory of frustrated quantum
  {H}eisenberg antiferromagnets},\ }\href
  {https://doi.org/10.1103/PhysRevB.41.9323} {\bibfield  {journal} {\bibinfo
  {journal} {Phys. Rev. B}\ }\textbf {\bibinfo {volume} {41}},\ \bibinfo
  {pages} {9323} (\bibinfo {year} {1990})}\BibitemShut {NoStop}%
\bibitem [{\citenamefont {Kotov}\ \emph {et~al.}(1998)\citenamefont {Kotov},
  \citenamefont {Sushkov}, \citenamefont {Weihong},\ and\ \citenamefont
  {Oitmaa}}]{Kotov1998}%
  \BibitemOpen
  \bibfield  {author} {\bibinfo {author} {\bibfnamefont {V.~N.}\ \bibnamefont
  {Kotov}}, \bibinfo {author} {\bibfnamefont {O.}~\bibnamefont {Sushkov}},
  \bibinfo {author} {\bibfnamefont {Z.}~\bibnamefont {Weihong}},\ and\ \bibinfo
  {author} {\bibfnamefont {J.}~\bibnamefont {Oitmaa}},\ }\bibfield  {title}
  {\bibinfo {title} {Novel approach to description of spin-liquid phases in
  low-dimensional quantum antiferromagnets},\ }\href
  {https://doi.org/10.1103/PhysRevLett.80.5790} {\bibfield  {journal} {\bibinfo
   {journal} {Phys. Rev. Lett.}\ }\textbf {\bibinfo {volume} {80}},\ \bibinfo
  {pages} {5790} (\bibinfo {year} {1998})}\BibitemShut {NoStop}%
\bibitem [{\citenamefont {Pearson}(1980)}]{Pearson1980}%
  \BibitemOpen
  \bibfield  {author} {\bibinfo {author} {\bibfnamefont {R.~B.}\ \bibnamefont
  {Pearson}},\ }\bibfield  {title} {\bibinfo {title} {Conjecture for the
  extended {P}otts model magnetic eigenvalue},\ }\href
  {https://doi.org/10.1103/PhysRevB.22.2579} {\bibfield  {journal} {\bibinfo
  {journal} {Phys. Rev. B}\ }\textbf {\bibinfo {volume} {22}},\ \bibinfo
  {pages} {2579} (\bibinfo {year} {1980})}\BibitemShut {NoStop}%
\bibitem [{\citenamefont {Kaufman}\ and\ \citenamefont
  {Onsager}(1949)}]{Kaufman1949}%
  \BibitemOpen
  \bibfield  {author} {\bibinfo {author} {\bibfnamefont {B.}~\bibnamefont
  {Kaufman}}\ and\ \bibinfo {author} {\bibfnamefont {L.}~\bibnamefont
  {Onsager}},\ }\bibfield  {title} {\bibinfo {title} {Crystal statistics. iii.
  short-range order in a binary {I}sing lattice},\ }\href
  {https://doi.org/10.1103/PhysRev.76.1244} {\bibfield  {journal} {\bibinfo
  {journal} {Phys. Rev.}\ }\textbf {\bibinfo {volume} {76}},\ \bibinfo {pages}
  {1244} (\bibinfo {year} {1949})}\BibitemShut {NoStop}%
\bibitem [{\citenamefont {Joshi}\ \emph {et~al.}(2015)\citenamefont {Joshi},
  \citenamefont {Coester}, \citenamefont {Schmidt},\ and\ \citenamefont
  {Vojta}}]{Joshi2015}%
  \BibitemOpen
  \bibfield  {author} {\bibinfo {author} {\bibfnamefont {D.~G.}\ \bibnamefont
  {Joshi}}, \bibinfo {author} {\bibfnamefont {K.}~\bibnamefont {Coester}},
  \bibinfo {author} {\bibfnamefont {K.~P.}\ \bibnamefont {Schmidt}},\ and\
  \bibinfo {author} {\bibfnamefont {M.}~\bibnamefont {Vojta}},\ }\bibfield
  {title} {\bibinfo {title} {Nonlinear bond-operator theory and $1/d$ expansion
  for coupled-dimer magnets. i. paramagnetic phase},\ }\href
  {https://doi.org/10.1103/PhysRevB.91.094404} {\bibfield  {journal} {\bibinfo
  {journal} {Phys. Rev. B}\ }\textbf {\bibinfo {volume} {91}},\ \bibinfo
  {pages} {094404} (\bibinfo {year} {2015})}\BibitemShut {NoStop}%
\end{thebibliography}%
\end{document}